\documentclass[fleqn,usenatbib]{mnras}

\usepackage{newtxtext,newtxmath}

\usepackage[T1]{fontenc}

\DeclareRobustCommand{\VAN}[3]{#2}
\let\VANthebibliography\thebibliography
\def\thebibliography{\DeclareRobustCommand{\VAN}[3]{##3}\VANthebibliography}

\usepackage{graphicx}	
\usepackage{amsmath}	

\usepackage[style=base,labelformat=simple]{subcaption} 

\title[X-Ray Microflaring Active Region]{NuSTAR observations of a repeatedly microflaring active region}

\author[K. Cooper et al.]
{Kristopher Cooper,$^{1}$\thanks{E-mail: k.cooper.2@research.gla.c.uk (KC)}
Iain G. Hannah,$^{1}$
Brian W. Grefenstette,$^{2}$
Lindsay Glesener,$^{3}$
S\"{a}m Krucker,$^{4,5}$
\newauthor
Hugh S. Hudson,$^{1,5}$
Stephen M. White,$^{6}$
David M. Smith,$^{7}$
and Jessie Duncan$^{3}$
\\
$^{1}$School of Physics \& Astronomy, University of Glasgow, University Avenue, Glasgow G12 8QQ, UK\\
$^{2}$Cahill Center for Astrophysics, California Institute of Technology, 1216 East California Boulevard, Pasadena, CA 91125, USA\\
$^{3}$School of Physics \& Astronomy, University of Minnesota Twin Cities, Minneapolis, MN 55455, USA\\
$^{4}$University of Applied Sciences and Arts Northwestern Switzerland, 5210 Windisch, Switzerland\\
$^{5}$Space Sciences Laboratory University of California, Berkeley, CA 94720, USA\\
$^{6}$Air Force Research Laboratory, Space Vehicles Directorate, Kirtland AFB, NM 87123, USA\\
$^{7}$Santa Cruz Institute of Particle Physics and Department of Physics, University of California, Santa Cruz, CA 95064, USA
}

\date{Accepted XXX. Received YYY; in original form ZZZ}

\pubyear{2015}
\begin{document}
\label{firstpage}
\pagerange{\pageref{firstpage}--\pageref{lastpage}}
\maketitle

\begin{abstract}
We investigate the spatial, temporal, and spectral properties of 10 microflares from AR12721 on~2018 September~9 and~10 observed in X-rays using the Nuclear Spectroscopic Telescope ARray (NuSTAR) and the Solar Dynamic Observatory's Atmospheric Imaging Assembly and Helioseismic and Magnetic Imager (SDO/AIA and HMI). We find GOES sub-A class equivalent microflare energies of 10$^{26}$--10$^{28}$~erg reaching temperatures up to 10~MK with consistent quiescent or hot active region core plasma temperatures of 3--4~MK. One microflare (SOL2018-09-09T10:33), with an equivalent GOES class of A0.1, has non-thermal HXR emission during its impulsive phase (of non-thermal power $\sim$7$\times$10$^{24}$~erg~s$^{-1}$) making it one of the faintest X-ray microflares to have direct evidence for accelerated electrons. In~4 of the~10 microflares, we find that the X-ray time profile matches fainter and more transient sources in the EUV, highlighting the need for observations sensitive to only the hottest material that reaches temperatures higher than those of the active region core ($>$5~MK). Evidence for corresponding photospheric magnetic flux cancellation/emergence present at the footpoints of 8 microflares is also observed.
\end{abstract}




\section{Introduction}

Active regions (ARs) are observed to sometimes produce repeated flares across decades of energies. These flaring processes are thought to be enabled by magnetic reconnection, which rapidly converts free magnetic energy into mass flows, particle acceleration, and plasma heating \citep{benz_flare_2017}.

Flares occur more frequently with decreasing energy and have a frequency consistent with a power-law magnitude distribution \citep[e.g.,][]{crosby_frequency_1993}. Therefore, microflares, the energetically weakest observed X-ray flares, are of particular interest as they may release more net energy into the solar atmosphere than their higher energy, less frequent counterparts. Microflares have energies about 10$^{26}$--10$^{28}$~erg and are identified to have $<$10$^{-6}$~W~m$^{-2}$ GOES (1--8~\AA{}) soft X-ray flux, labelled as B, A, and sub-A class flares \citep{lin_solar_1984, fletcher_observational_2011, hannah_microflares_2011}. Sub-A class microflares are not reliably detected by GOES but sub-A level events have been observed by more sensitive full-disk X-ray spectrometers identified to be located in the quiet Sun \citep{sylwester_sphinx_2012, vadawale_observations_2021} and ARs \citep{gburek_soft_2011, vadawale_observations_2021-1}. However, a GOES equivalent class for sub-A class microflares can be calculated from their temperature and emission measure.

Even weaker flares with energies about 10$^{24}$~erg (nanoflares) are proposed to take place everywhere, not just localised to ARs \citep{parker_nanoflares_1988}. If the frequency distribution has a negative power-law $>$2 then weaker flares could provide a majority fraction of the total power heating the corona. However, this relies on similar properties and processes, such as non-thermal energy release mechanisms, being present as the energy of the flare scales down \citep{hudson_solar_1991}.

X-ray microflares have been studied previously in great detail with instruments including the Reuven Ramaty High-Energy Solar Spectroscopic Imager \citep[RHESSI;][]{lin_reuven_2002} and the Nuclear Spectroscopic Telescope ARray \citep[NuSTAR;][]{harrison_nuclear_2013}. Previous studies have observed and quantified evidence of hard X-ray (HXR) non-thermal emission produced during microflares suggesting that similar physics does indeed operate across decades of flare energies; however, the physical size of the microflare is not necessarily scaled with its energy release \citep{christe_rhessi_2008, hannah_rhessi_2008, glesener_accelerated_2020}. Further work on low A-class microflares has also found that HXR emission commonly peaks before lower energy emission, a sign of hotter emission being present in the earlier stages of the flaring process or indicative of non-thermal processes \citep{duncan_nustar_2021}. Therefore, HXR sensitivity at these small scales is crucial to further our understanding of the mechanisms in weak solar flares and, thus, solar atmospheric heating.

NuSTAR is an astrophysical HXR focusing optics imaging spectrometer that is sensitive to photon energies between 2.5--79~keV and capable of observing the Sun \citep{grefenstette_first_2016, hannah_first_2016}. NuSTAR uses \ion{Wolter-}{i} type optics to focus X-rays onto two focal plane modules (FPMA and FPMB). Each FPM has a field-of-view of 12\arcmin$\times$12\arcmin with small gaps between the four detector chips. The optics' point spread function (PSF) has a full-width half maximum of 18\arcsec{} and a half-power diameter of 58\arcsec{} \citep{harrison_nuclear_2013}. Each detected photon is processed over 2.5~ms per FPM where no other triggering event can be recorded. The time spent open to detection is termed the livetime. Even with small microflares (A-class and smaller) the NuSTAR livetime is low ($<$16\%), hindering the detection of photons from relatively weakly emitting higher energy sources. 

NuSTAR has observed several B, A, and equivalent sub-A class AR microflares of energies from 10$^{28}$~erg down to 10$^{26}$~erg \citep{glesener_nustar_2017, wright_microflare_2017, hannah_joint_2019, cooper_nustar_2020,duncan_nustar_2021}. Quiet Sun brightenings outside ARs have also been observed with thermal energies of 10$^{26}$~erg \citep{kuhar_nustar_2018}. Observations of such small events are possible due to NuSTAR's sensitivity during the solar minimum between cycle 24 and 25. Non-thermal emission has also been directly observed with NuSTAR's focusing optics imaging spectroscopy in a GOES class A5.7 microflare \citep{glesener_accelerated_2020} with other studies finding a non-thermal power source consistent with the microflares under investigation \citep{wright_microflare_2017, cooper_nustar_2020,duncan_nustar_2021}.\footnote{An overview of NuSTAR solar observations is available at \url{https://ianan.github.io/nsigh_all/}}
   
In this paper we present observations of all identified microflares observed by NuSTAR from AR12721 on 2018 September~9--10. In Section~\ref{sec:overview} we present an overview of the whole NuSTAR campaign across both days of observations and discuss the broad similarities in each microflare's analysis. We then describe each identified microflare's X-ray and extreme-ultraviolet (EUV) spatial and temporal data using 12~s cadence images from the Solar Dynamics Observatory's Atmospheric Imaging Assembly \citep[SDO/AIA;][]{lemen_atmospheric_2012}, along with the corresponding X-ray spectra, in Section~\ref{sec:temporal and spatial and spectral}. Using 45~s cadence data from the Solar Dynamics Observatory's Heliospheric and Magnetic Imager \citep[SDO/HMI;][]{schou_design_2012}, the presence of mixed polarity photospheric magnetic flux at, or close to, the apparent footpoints of eight microflares is discussed in Section~\ref{sec:hmi}.

\section{NuSTAR observations: 2018 September 9--10} \label{sec:overview}

NuSTAR observed the Sun on 2018 September 9 and 10 performing six hour-long dwells initially set to target a region previously investigated by the FOXSI-3 sounding rocket \citep{musset_ghost-ray_2019} on September 7. However, AR12721 appeared on September 8 and subsequently dominated NuSTAR's field of view (FOV). 

Furthermore, due to this unexpected AR appearance, the pointing of NuSTAR was not optimal with some microflares obscured by the detector chip-gaps. Therefore, analysis of some microflares must rely on only one FPM or, in a few unfortunate situations, could not be investigated at all since neither FPM was suitable.

\begin{figure*}
	\centering
	
    \begin{subfigure}{1.4\columnwidth}
    \includegraphics[width=\columnwidth,height=0.65\linewidth]{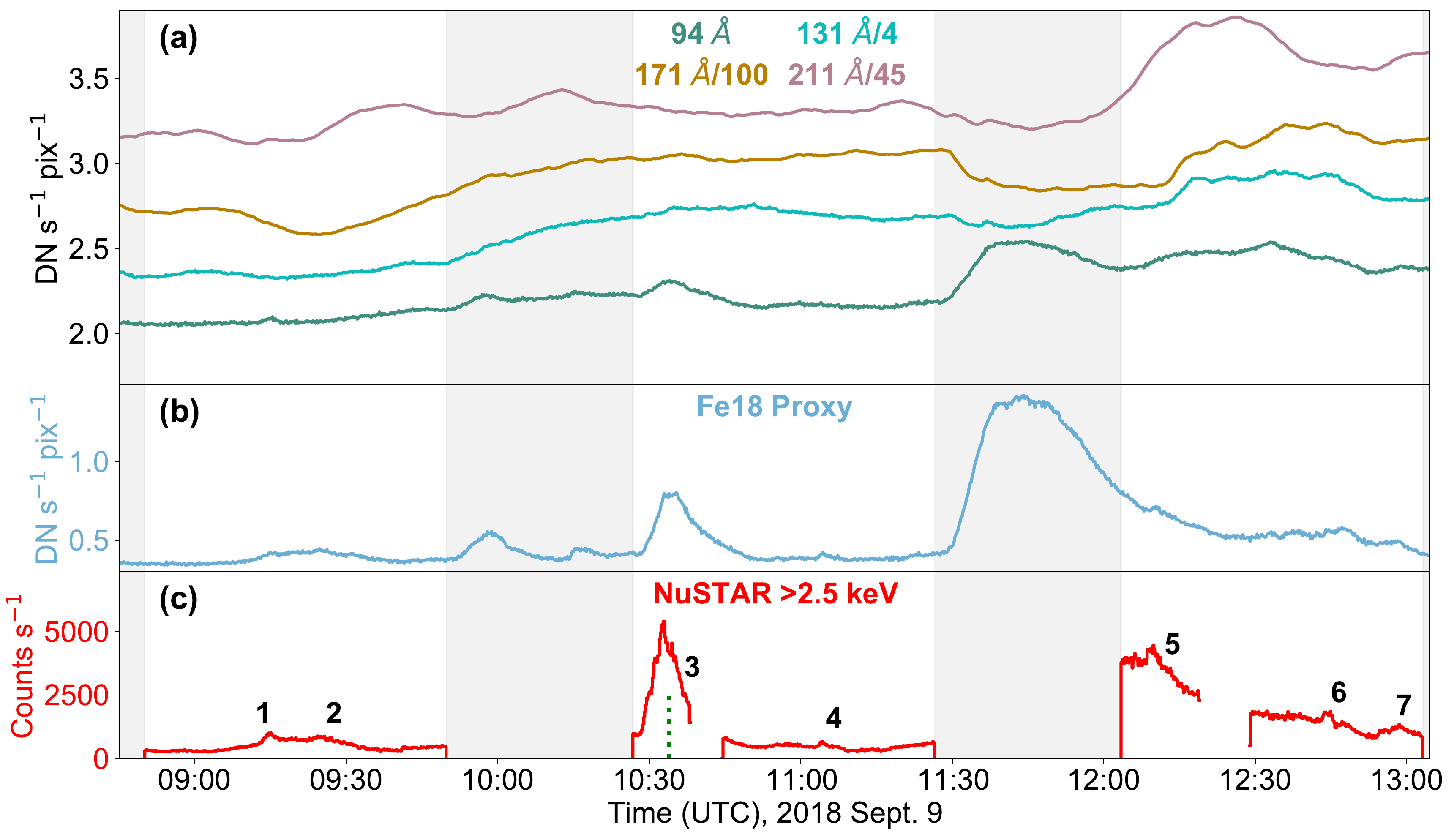}
    \end{subfigure}
    \begin{minipage}{0.67\columnwidth}
    \begin{subfigure}{\columnwidth}
    \includegraphics[width=\linewidth]{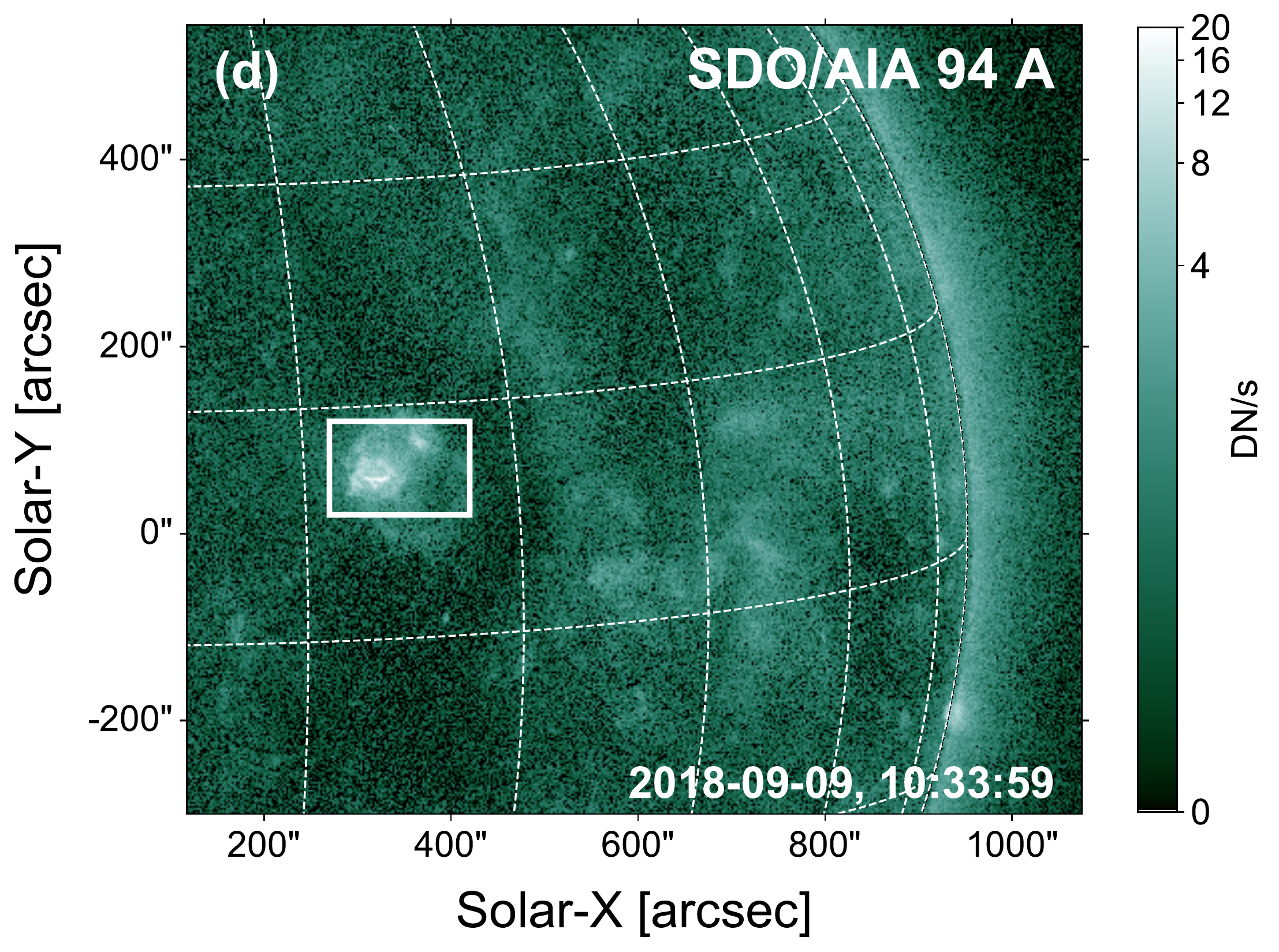}
    \end{subfigure}
    \begin{subfigure}{\columnwidth}
    \includegraphics[width=\linewidth]{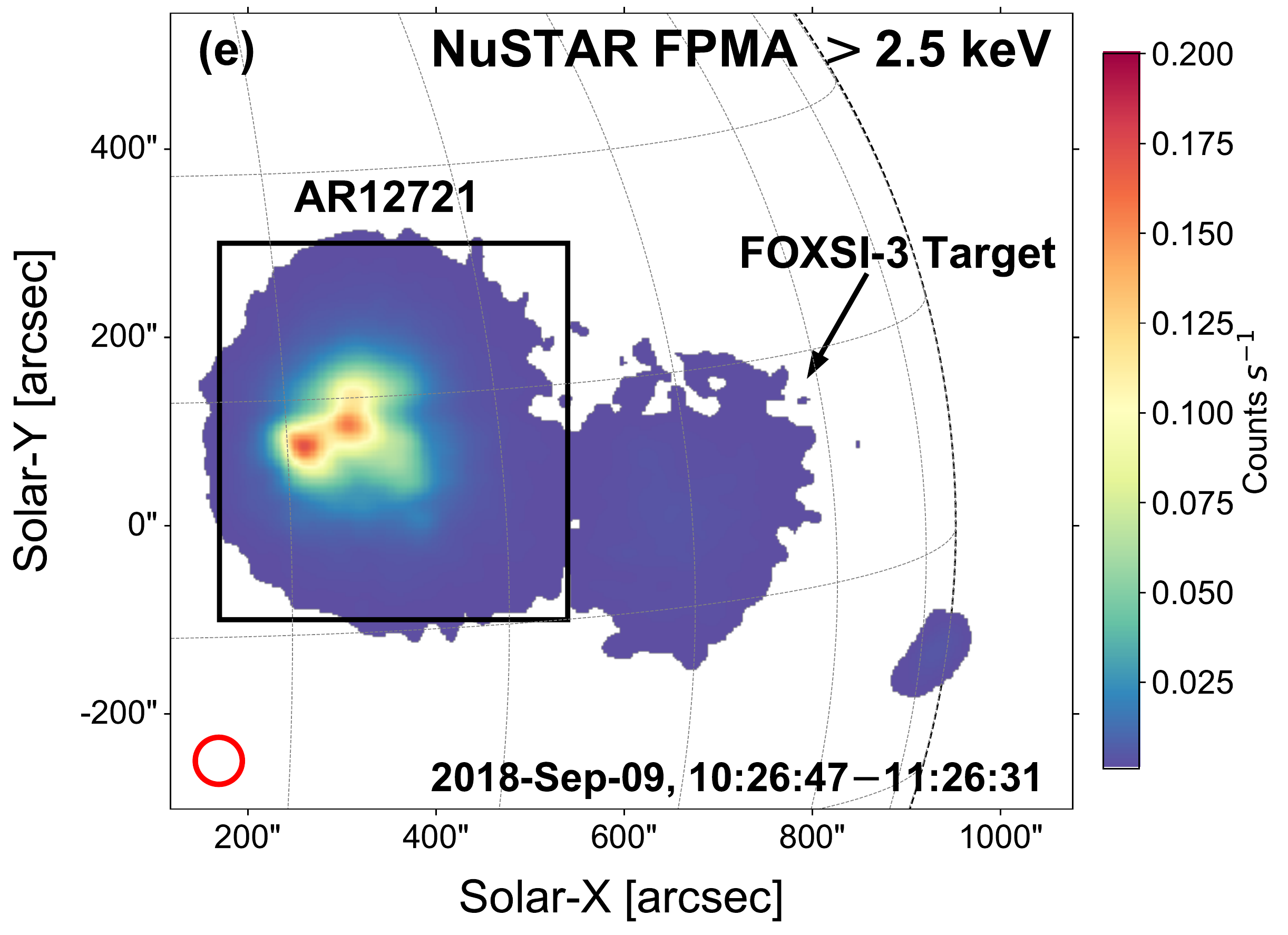}
    \end{subfigure}
    \end{minipage}
    
    \begin{subfigure}{1.4\columnwidth}
    \includegraphics[width=\columnwidth, height=0.65\linewidth]{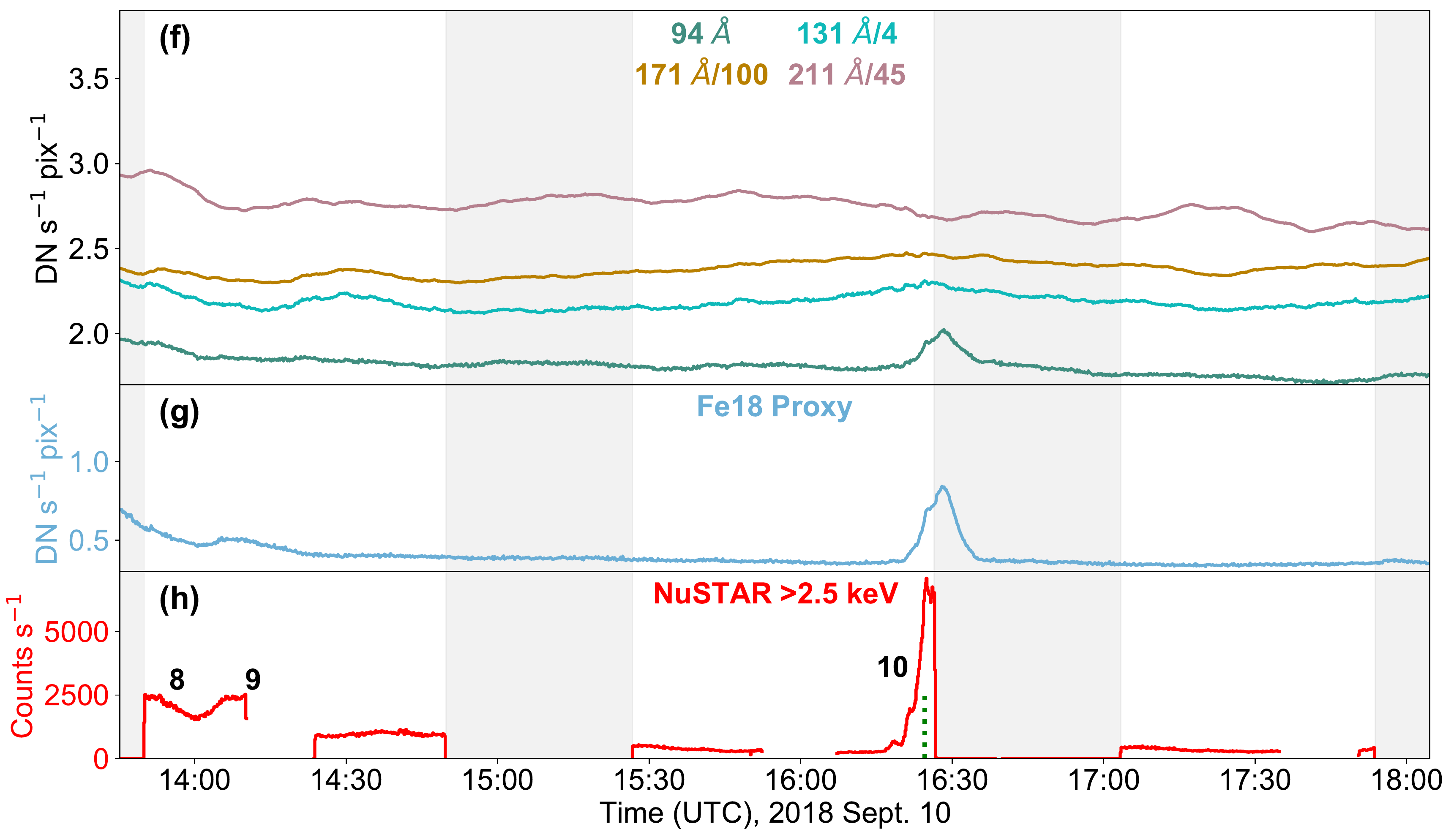}
    \end{subfigure}
    \begin{minipage}{0.67\columnwidth}
    \begin{subfigure}{\columnwidth}
    \includegraphics[width=\linewidth]{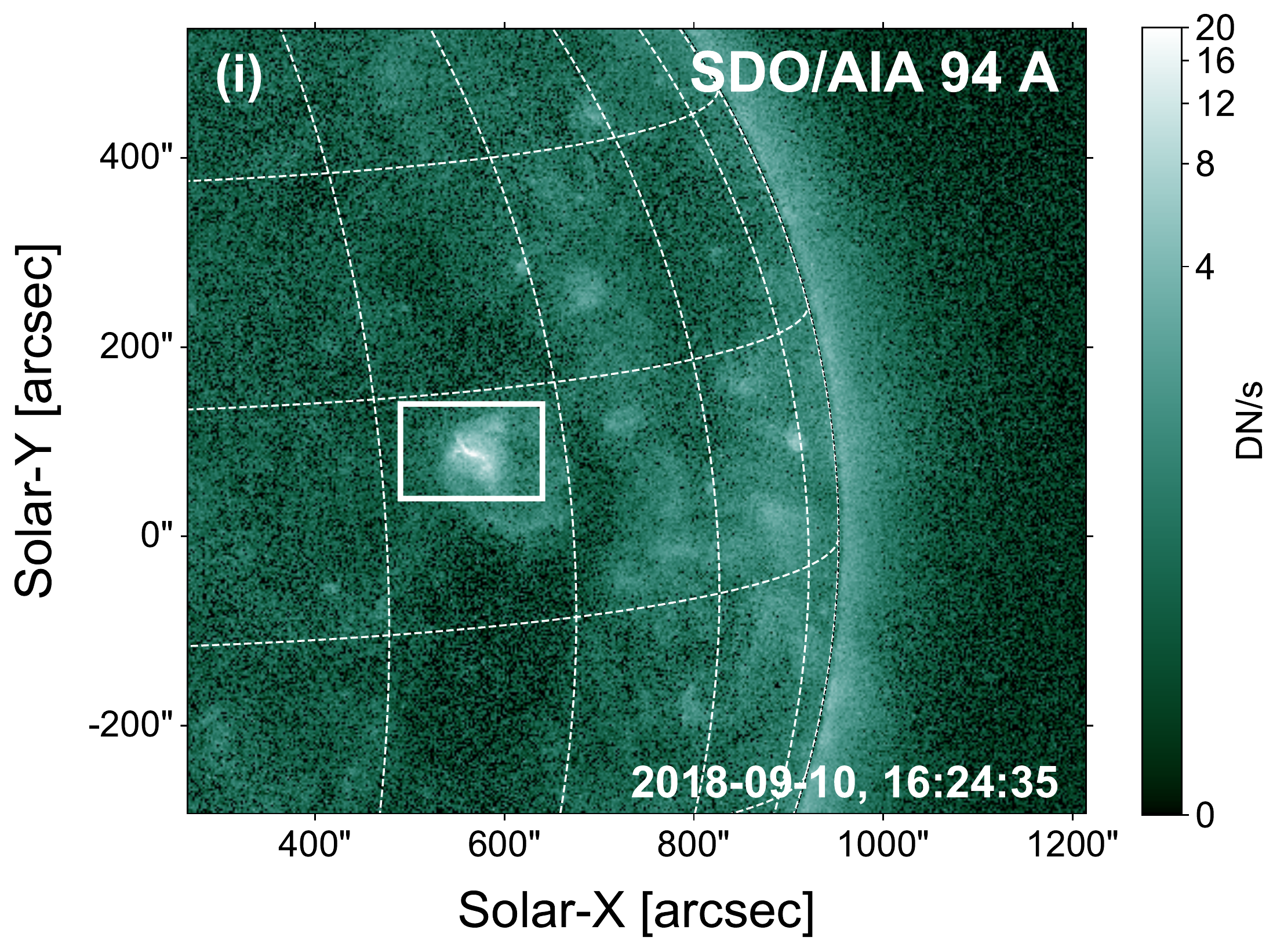}
    \end{subfigure}
    \begin{subfigure}{\columnwidth}
    \includegraphics[width=\linewidth]{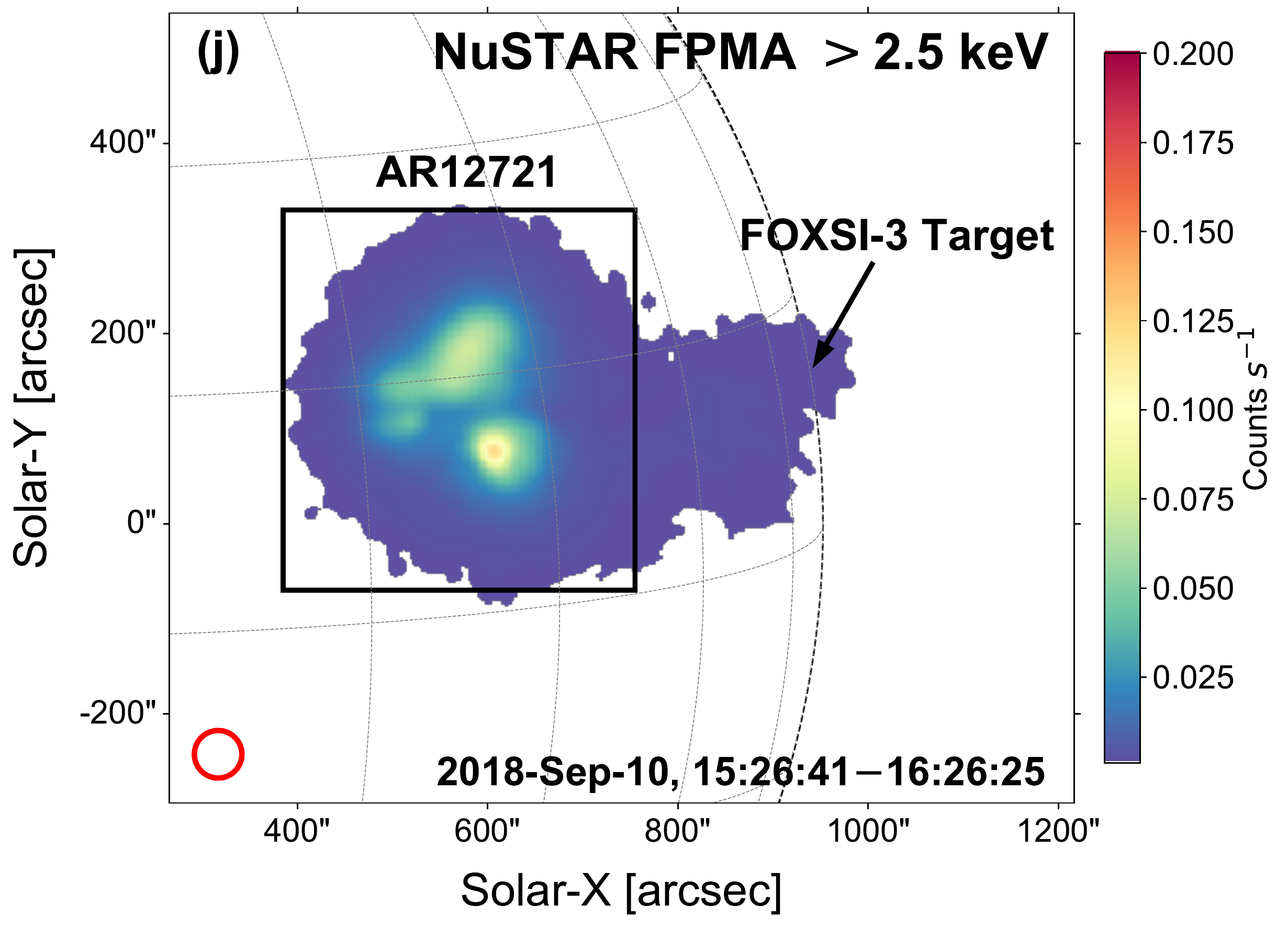}
    \end{subfigure}
    \end{minipage}
    \caption{SDO/AIA, including \ion{Fe}{xviii} proxy, and NuSTAR lightcurves from AR12721 on 2018 September 9 (panels~a--c) and 10 (panels~f--h). The areas used to obtain the NuSTAR and SDO/AIA time profiles cover the full AR over each day shown by the box regions in panels~d,~e and~i,~j for September~9 and~10, respectively. The SDO/AIA 94~\AA{} images (panel~d and~i) are taken from the times indicated by the time stamps and vertical, dotted green lines in panel~c and~h while each FOV NuSTAR image (panel~e and~j) shows the integrated FPMA $>$2.5~keV emission over the second orbit on each day (orbit~2 and~4). The red circles with diameters of 50\arcsec{} in panels~e and~j indicate the minimum size of region used to produce spectral profiles for each microflare. The shaded regions indicate NuSTAR's eclipse with any gaps in the data outside of these grey periods being due to SAA passage. The same y-axis limits are used between the top and bottom lightcurve panels. The channel of the lightcurve emission is displayed at the top of each time profile with a scaling factor, if required. NuSTAR lightcurves and images are livetime corrected and the numbers from 1--10 indicate the identified microflares.}
    \label{fig:lightcurves_9th10th}
\end{figure*}

In order to help compare the X-ray data with the EUV SDO/AIA channels, and provide higher resolution spatial context for the hotter material, we calculate an \ion{Fe}{xviii} proxy \citep{del_zanna_multi-thermal_2013} from a linear combination of the degradation-corrected SDO/AIA 94~\AA{}, 171~\AA{}, and 211~\AA{} channels. The \ion{Fe}{xviii} proxy is sensitive to material between 4--10~MK which are temperatures expected to be present in flaring coronal plasma \citep{odwyer_sdoaia_2010, lemen_atmospheric_2012,warren_systematic_2012}. In addition, the SDO/AIA 131~\AA{} channel is sensitive to material $>$10~MK; however, Figure~\ref{fig:lightcurves_9th10th} shows its evolution is similar to lower temperature SDO/AIA channels and does not match well to the NuSTAR time profile. Therefore, most of the hotter material in these data is likely constrained below 10~MK. Therefore, the \ion{Fe}{xviii} synthetic channel is an appropriate candidate to corroborate the hotter thermally emitting material that NuSTAR observes from microflares \citep{hannah_first_2016, wright_microflare_2017, hannah_joint_2019, glesener_accelerated_2020,duncan_nustar_2021}.

Figure~\ref{fig:lightcurves_9th10th} (panels~a--c and~f--h) shows the lightcurves from four SDO/AIA channels, the SDO/AIA \ion{Fe}{xviii} synthetic channel, and NuSTAR X-ray emission $>$2.5~keV from both FPMA\&B over all six NuSTAR orbits. Although one FPM may have provided better quality data than the other at certain time intervals, both generally provide corroborative qualitative agreement with each other. The boxed areas in panel~d and~i indicate the areas used to produce the SDO/AIA lightcurves while panel~e and~j show the size of region used for the NuSTAR time profiles for each day. From the NuSTAR FPMA FOV images (panel~e and~j) it is clear that AR12721 dominated over the initially targeted FOXSI-3 region. 

To identify features of interest for analysis from AR12721 time profiles were produced over the six dwell times from four SDO/AIA channels, the \ion{Fe}{xviii} proxy, and NuSTAR grade 0 (single pixel) FPMA+B X-ray counts $>$2.5~keV (Figure~\ref{fig:lightcurves_9th10th}). This reveals the presence of 10 microflares over the first 5 orbits (labelled 1--10) in X-rays with only weak correlations in the the native SDO/AIA channels. Event times were identified with NuSTAR and only then corroborated with \ion{Fe}{xviii}. The last dwell did not appear to show the presence of activity but may be useful in studying quiescent AR emission. 

The numerous microflares NuSTAR observed originating from AR12721 all varied in spatial complexity as seen in \ion{Fe}{xviii}. The 10 identified events were analysed providing varying examples of clear, whole or partial, microflare time profiles and indications of loop heating. One such event, microflare~4 in Figure~\ref{fig:lightcurves_9th10th} (panel c), was previously found to be the weakest X-ray microflare in literature \citep{cooper_nustar_2020}. 

\section{AR12721 X-ray Microflares} \label{sec:temporal and spatial and spectral}

NuSTAR images were created by spatially binning detected counts from the event lists integrated over the microflare time, or a sub-range of it. We then deconvolve the X-ray emission map with the PSF using the Richardson-Lucy method \citep{richardson_bayesian-based_1972}. NuSTAR's pointing is controlled by star trackers, whose camera head unit (CHU) combinations change throughout each orbit and so the deconvolved X-ray image is then co-aligned with an SDO/AIA \ion{Fe}{xviii} image for a given CHU state. We investigate the NuSTAR emission in two energy ranges, 2.5--4~keV and 4--10~keV, as the X-ray emission $<$4~keV tends to closely follow the \ion{Fe}{xviii} emission and each microflare generally shows at least a $\sim$100\% increase in X-ray emission $>$4~keV when a microflare occurs. There was no significant emission $>$10~keV.

The number of iterations for deconvolution was~100 and~50 for the 2.5--4~keV and 4--10~keV energy channels, respectively, in the two brightest microflares (3 and~10). For the other, weaker microflares the iterations were~80 and~40 for the same energy bands. Iteration numbers were investigated from 25 to 200 in increments of 25 for each energy range and it was found that the general shape did not change for any event, it only becomes more compact and pronounced with increasing number of iterations. In addition, NuSTAR's PSF profile changes shape with radial distance from the optical axis (OA) and orientation with azimuthal angle \citep{madsen_calibration_2015}. We take these parameters into account producing contours that match slightly better with the features seen in \ion{Fe}{xviii}; however, no additional or different structures are revealed.

The NuSTAR spectra were fitted with XSPEC \citep{arnaud_astronomical_1996}, making use of Cash statistics \citep{cash_parameter_1979} to aid the low count regimes. APEC thermal models and broken power-law models were used to probe the thermal and non-thermal nature of each microflare's spectrum with 1-$\sigma$ equivalent errors produced through MCMC analysis for all fitted and derived parameters. Coronal abundances were assumed \citep{feldman_potential_1992} and all spectra were produced by integrating over a circular region $>$50\arcsec{} in diameter centered on the brightest emission. The region for the microflare is made large enough to account for NuSTAR's PSF where the inclusion of any surrounding AR material would only add to the hot AR core (3--4~MK) component; a 50\arcsec{} diameter circle is shown in Figure~\ref{fig:lightcurves_9th10th} (panels~e and~j) for scale. A similar statement can be made with regards to the time ranges chosen to produce the spectra; long enough to provide sufficient signal-to-noise but not too long to mix together different stages of the temporal evolution. No gain correction was required as the livetime was relatively large, compared to the microflares studied by \cite{duncan_nustar_2021}.

\subsection{Orbit 1: Microflare 1 and 2} \label{sec:orbit 1}

\begin{figure*}
	\centering
	\includegraphics[width=2.08\columnwidth]{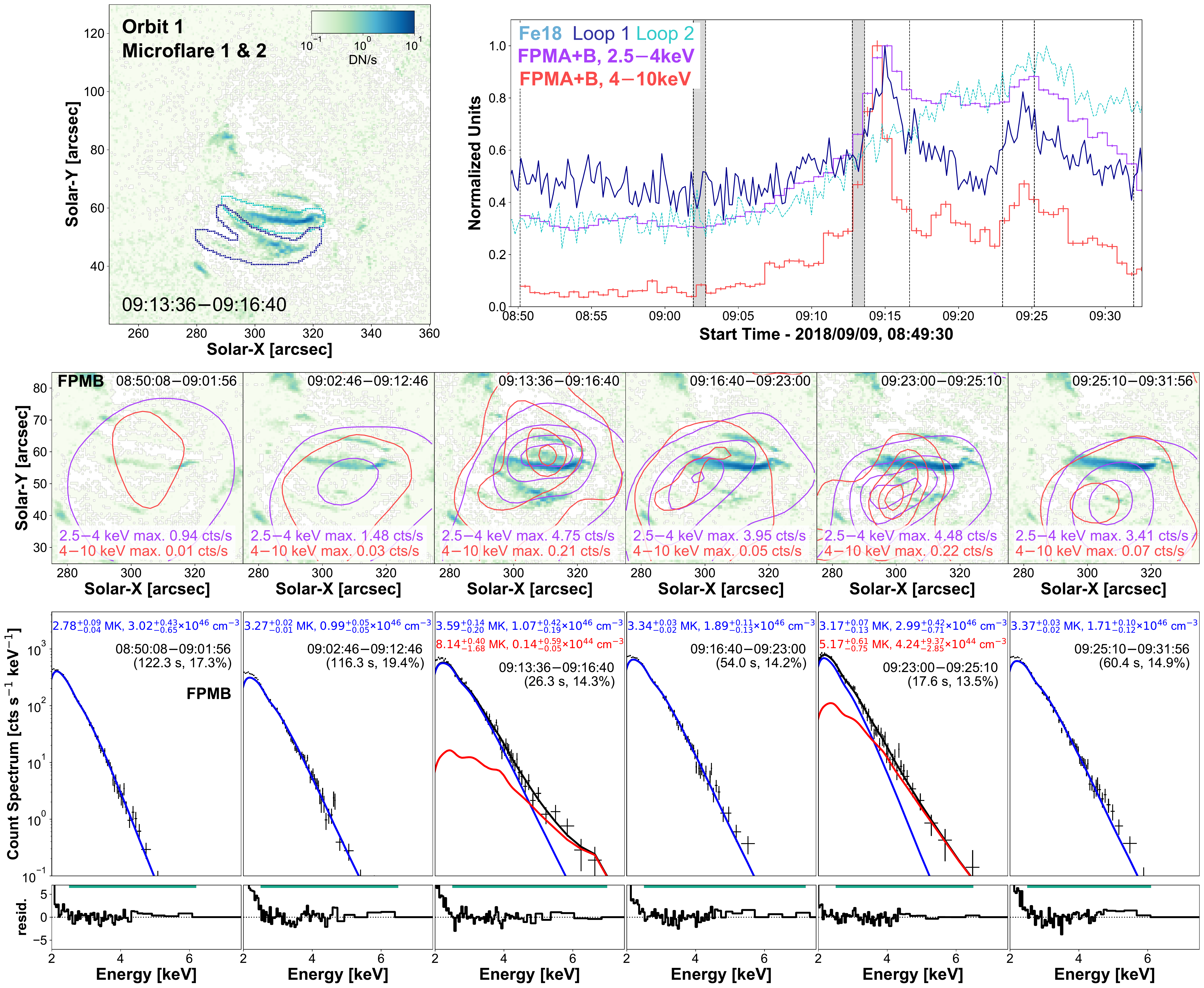}
    \caption{SDO/AIA \ion{Fe}{xviii} image integrated over microflare~1's time (top left panel) with sources identified to have corresponding time series to NuSTAR outlined in navy and cyan. Maximum normalised time profiles from the two loops visible in \ion{Fe}{xviii} (navy and cyan) and two NuSTAR FPMA+B energy ranges (2.5--4~keV: purple, 4--10~keV: red, 40~s binning) are plotted (top right panel). The area used to produce the \ion{Fe}{xviii} lightcurve is shown in the top left panel with the appropriate coloured contour, whereas the NuSTAR lightcurves are integrated over the full AR. The vertical dashed lines indicate a pre-flare time (08:50:08--09:01:56~UTC), a rise time (09:02:46--09:12:46~UTC), microflare~1's time (09:13:36--09:16:40~UTC), and microflare~2's time (09:23:00--09:25:10~UTC) with the dark grey regions indicating times that are not used due to a noticeable shift in source location via CHU state changes. A single shift correction was applied to each CHU combination. NuSTAR contours (middle row) are created over the time shown at the top right of each panel and indicated by the vertical dashed lines in the lightcurve panel. The FPMB 2.5--4~keV and 4--10~keV emission is overlain on the average \ion{Fe}{xviii} image and show the 2, 20, 50, and 80\% of the maximum emission across all time ranges for each energy range (2.5--4~keV (purple): 4.75~counts~s$^{-1}$, 4--10~keV (red): 0.22~counts~s$^{-1}$). The X-ray spectra over the same time ranges are shown directly below in the bottom row with the fitting range indicated by the horizontal green line above the residuals. The temperature, emission measure, time range, and livetime of each fit is also shown. Errors presented in the spectral plots are 1-sigma equivalent and were obtained through MCMC analysis. The y-range displayed in the top left panel is used for all similar panels in other microflare figures for spatial context. It should be noted that the higher/lower uncertainty on temperature corresponds to the lower/higher emission measure uncertainty.}
    \label{fig:m1and2}
\end{figure*}

During orbit~1 we identify two microflares (labelled~1 and~2) producing raised X-ray emission at 09:15~UTC and 09:25~UTC (Figure~\ref{fig:m1and2}). Two sets of loops appear spatially resolved in \ion{Fe}{xviii} (Figure~\ref{fig:m1and2}, top left panel) with the navy contour identifying the loops that show a very similar impulsive profile to microflare~1 and~2 (09:15~UTC and 09:25~UTC) as seen from the X-ray lightcurve (Figure~\ref{fig:m1and2}, top right panel, navy). Microflare~2's X-rays also coincide with the peak of the slowly varying EUV emission from the cyan contour. 

When plotting X-ray emission contours on top of the average \ion{Fe}{xviii} emission for each defined time range we find that NuSTAR does not indicate multiple sources (Figure~\ref{fig:m1and2}, middle row), even when the EUV lightcurves suggest two resolved sites are likely to be contributing emission. NuSTAR FPMB images were co-aligned with the \ion{Fe}{xviii} emission using only a single shift per CHU state. This could indicate that the movement of the source between 09:13:36--09:31:56~UTC is real; however, this movement is less pronounced in X-ray images that are not deconvolved. Therefore, this perceived drift could be caused by the deconvolution process instead. This movement meant a single shift would not align the X-ray and EUV source for the full time between 09:13:36--09:31:56~UTC.

Due to the resolution of the X-ray images there is ambiguity as to which EUV loop the microflares originate. It is possible that the NuSTAR emission from microflare~2 may come from either or both loops as there are corresponding peaks from both loops during microflare~2's time. This could indicate that both loops are physically connected through some means. Therefore, the processes in one loop may be able to affect the material present in the other or a third feature could be driving the increased emission in both loops.

Fitting the spectra obtained over the time ranges shown in Figure~\ref{fig:m1and2} (bottom row) with APEC thermal models we find that the pre-flare phase starts at a temperatures of $\sim$2.8~MK and emission measure of 3.0$\times$10$^{46}$~cm$^{-3}$. During the rise time some pre-flare material is heated to 3.3~MK and 1.0$\times$10$^{46}$~cm$^{-3}$ before continuing to be enhanced to 3.6~MK during microflare~1. However, during microflare~1 an excess appears indicating that this microflare reaches temperatures of 8.1~MK with a small emission measure of 1.4$\times$10$^{43}$~cm$^{-3}$. In reality, plasma will be heated to a continua of temperatures during microflare~1 (09:13:36--09:16:40~UTC) which could explain the rise in temperature for the non-excess thermal model for this time. After microflare~1 the excess disappears and the pre-flare emission continues to be enhanced in terms of emission measure, up to 1.9$\times$10$^{46}$~cm$^{-3}$, as more material is heated $>$3~MK. 

Then, during microflare~2 (09:23:00--09:25:10~UTC), an excess above an isothermal model appears again indicating that temperatures of $\sim$5~MK are reached. The decay (09:25:10--09:31:56~UTC) then shows that the plasma drops back to a similar state to that before microflare~2. All times investigated during orbit~1 show the presence of $\sim$3--4~MK plasma, consistent with previous AR temperatures found by NuSTAR \citep{wright_microflare_2017,glesener_nustar_2017,hannah_joint_2019, cooper_nustar_2020}.

By estimating the volume of both sets of loops observed in \ion{Fe}{xviii} we then obtain an estimate for the instantaneous thermal energy released for both microflare~1 and~2. The loop volumes (V) are calculated by modelling the navy contour structure as two loops and the cyan contour structure as one loop with a half-torus geometry (Figure~\ref{fig:m1and2} top left panel). This gives a volume of 2.2$\times$10$^{26}$~cm$^{3}$ and 8.4$\times$10$^{25}$~cm$^{3}$ for the navy and cyan contour loops, respectively.

Using Equation~3 in \cite{hannah_rhessi_2008} in conjunction with the excess/microflare temperatures and emission measures, and assuming the microflares occur in the navy contour loop as suggested by the time profiles, we find instantaneous thermal energies of 1.87$^{+1.51}_{-0.26}\times$10$^{26}$~erg and 6.53$^{+3.47}_{-2.36}\times$10$^{26}$~erg for microflare~1 and~2, respectively, where the volume filling factor is assumed to be 1. If we consider that microflare~2's energy release involved both the navy and cyan loops it would become 7.69$^{+4.11}_{-2.77}\times$10$^{26}$~erg. We also calculate the GOES equivalent class for each event, with the excess/microflare temperatures and emission measures, via the \verb|goes_flux49.pro|\footnote{\url{https://hesperia.gsfc.nasa.gov/ssw/gen/idl/synoptic/goes/goes_flux49.pro}} IDL routine with default coronal abundances which calls \verb|CHIANTI V7.1| \citep{dere_chianti_1997,landi_chiantiatomic_2013}. Microflare~1 and~2 was calculated to be GOES sub-A-class equivalent with classes of A0.001 and A0.01, respectively. Spectral fit parameters for all but microflare~3 and~10 are displayed in Table~\ref{tab:xspec fits}.

\subsection{Orbit 2: Microflare 3 and 4} \label{sec:orbit 2}

\begin{figure*}
	\centering
	\includegraphics[width=2\columnwidth]{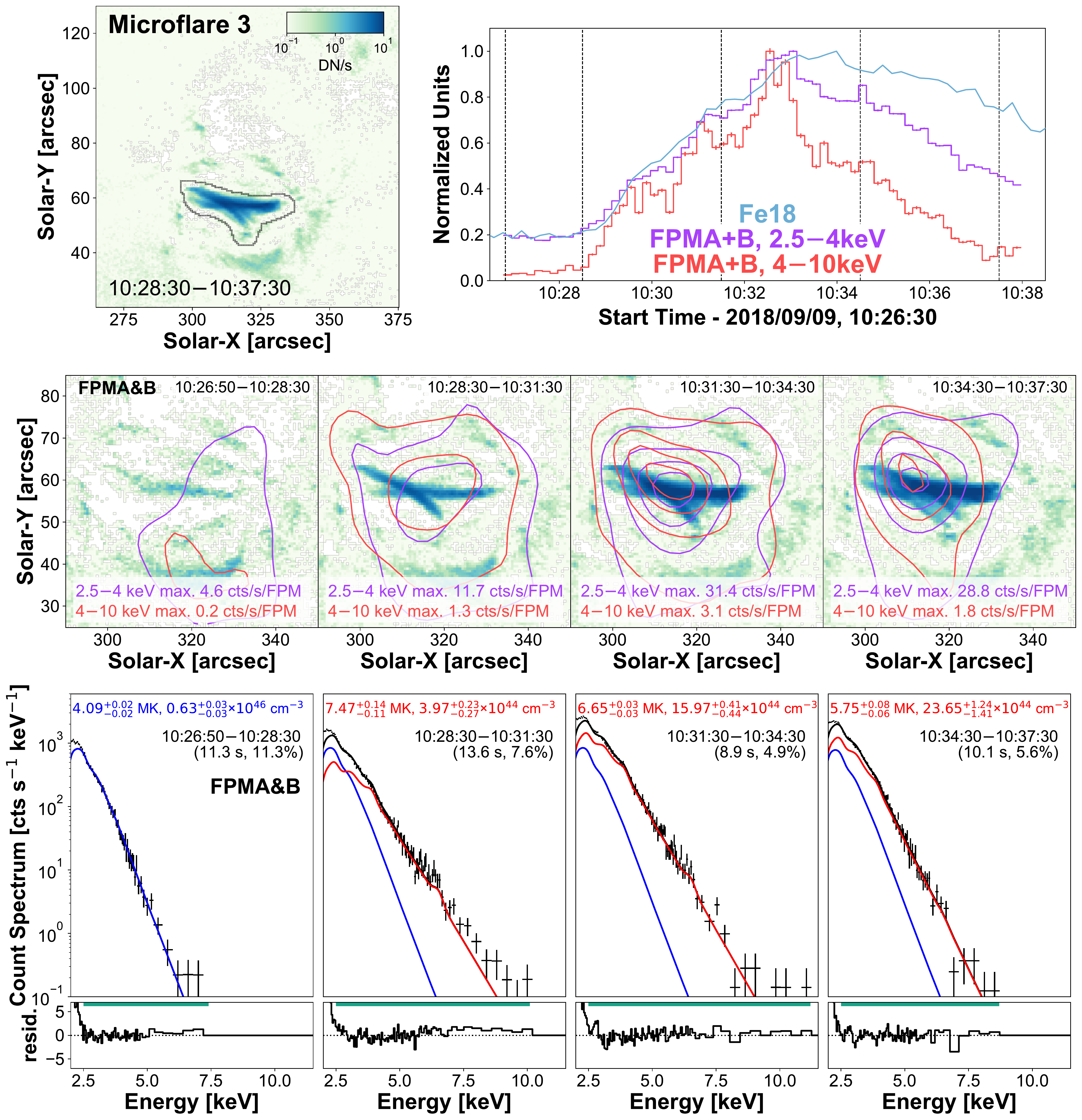}
    \caption{The \ion{Fe}{xviii} emission integrated over microflare~3 (top left panel). The black contours show the region used to determine the \ion{Fe}{xviii} time profile (blue, top right panel). NuSTAR FPMA+B 2.5--4~keV (purple) and 4--10~keV (red) emission is integrated over the full AR with 10~s binning. The middle row shows NuSTAR FPMA\&B combined contours, integrated over the time ranges indicated in the lightcurve plot by vertical dashed lines where the contour levels are 2, 20, 50, and 80\% of the maximum emission across all time ranges for each energy range (2.5--4~keV (purple): 31.4~counts~s$^{-1}$~FPM$^{-1}$, 4--10~keV (red): 3.1~counts~s$^{-1}$~FPM$^{-1}$). The \ion{Fe}{xviii} image is the average emission over the respective time range. Corresponding spectral fits of the four times are shown in the bottom row, with temperature, emission measure, time range, and livetime displayed. The pre-flare thermal parameters (bottom left panel, blue) were a fixed components in the microflare times.}
    \label{fig:m3}
\end{figure*}

Microflare~3 is one of the brightest X-ray events observed throughout the two-day period. Although the time profile of microflare~3 appears relatively simple (Figure~\ref{fig:m3}, top right panel) the image (top left panel) shows a network of loops being heated. Due to the duration of microflare~3 it is easily broken down into three time ranges over a period of 9~minutes---a rise, peak, and decay time---with the addition of a quiescent pre-flare time (Figure~\ref{fig:m3}, top right panel).

The pre-flare time (10:26:50--10:28:30~UTC) does not provide any concentrated emission where microflare~3 takes place, suggesting that the X-ray emission is dominated by slowly varying, non-flaring AR emission (Figure~\ref{fig:m3}, middle row, left panel). This is supported when the simultaneously fitted FPMA and B spectrum from this time (Figure~\ref{fig:m3}, bottom row, left panel) is consistent with typical hot AR core temperatures at $\sim$4~MK \citep{warren_systematic_2012}.

Both FPMA and B are usable throughout microflare~3 and are both deconvolved then co-aligned with \ion{Fe}{xviii} separately before being combined (Figure~\ref{fig:m3}, middle row). A single shift, determined from the peak of microflare~3 (10:31:30--10:34:30~UTC), is applied across all four times and both energy ranges. The contours of the different energy ranges have the same shape during the microflare with no significant difference in centroid location. Throughout the microflare the centroids appear to move slightly to the left.

Microflare~3's pre-flare thermal model with a temperature 4.1~MK and emission measure of 6.3$\times$10$^{46}$~cm$^{-3}$ was a fixed component in the rise, peak, and decay spectra. Spectral fitting (Figure~\ref{fig:m3}, bottom row) indicates that the initial phase of the microflare is the hottest with a temperature of 7.5~MK and emission measure of 4.0$\times$10$^{44}$~cm$^{-3}$. 

The rise phase plasma then cools slightly to 6.7~MK at the peak then to 5.8~MK during the decay while increasing the emission measure to 16.0$\times$10$^{44}$~cm$^{-3}$ then finally to 23.7$\times$10$^{44}$~cm$^{-3}$, respectively. Therefore, as the microflare progresses from the rise to the peak and then decay phase chromospheric evaporation takes place continually expanding heated chromospheric plasma into the coronal loops \citep{fletcher_observational_2011}.

We use the temperatures and emission measures obtained from spectral fitting the microflare excess plasma to quantitatively compare with the SDO/AIA \ion{Fe}{xviii} proxy channel. We find good agreement between the emission that is modelled from the NuSTAR X-ray spectrum and that observed in the \ion{Fe}{xviii} pre-flare subtracted emission when folding the microflare excess models through the SDO/AIA \ion{Fe}{xviii} temperature response (NuSTAR sees $\sim$42\% during the rise, $\sim$65\% at peak, and $\sim$75\% during decay).

Microflare~3's rise time spectrum (10:28:30--10:31:30~UTC) shows an excess above the total model fit $>$7~keV suggesting another model component is needed to represent the observed emission. Fitting an additional APEC thermal model we find an unphysically high temperature ($\sim$95~MK) is required; therefore, we fit a power-law model (representing non-thermal emission) to characterise the excess.

\begin{figure}
	\centering
	\includegraphics[width=0.9\columnwidth]{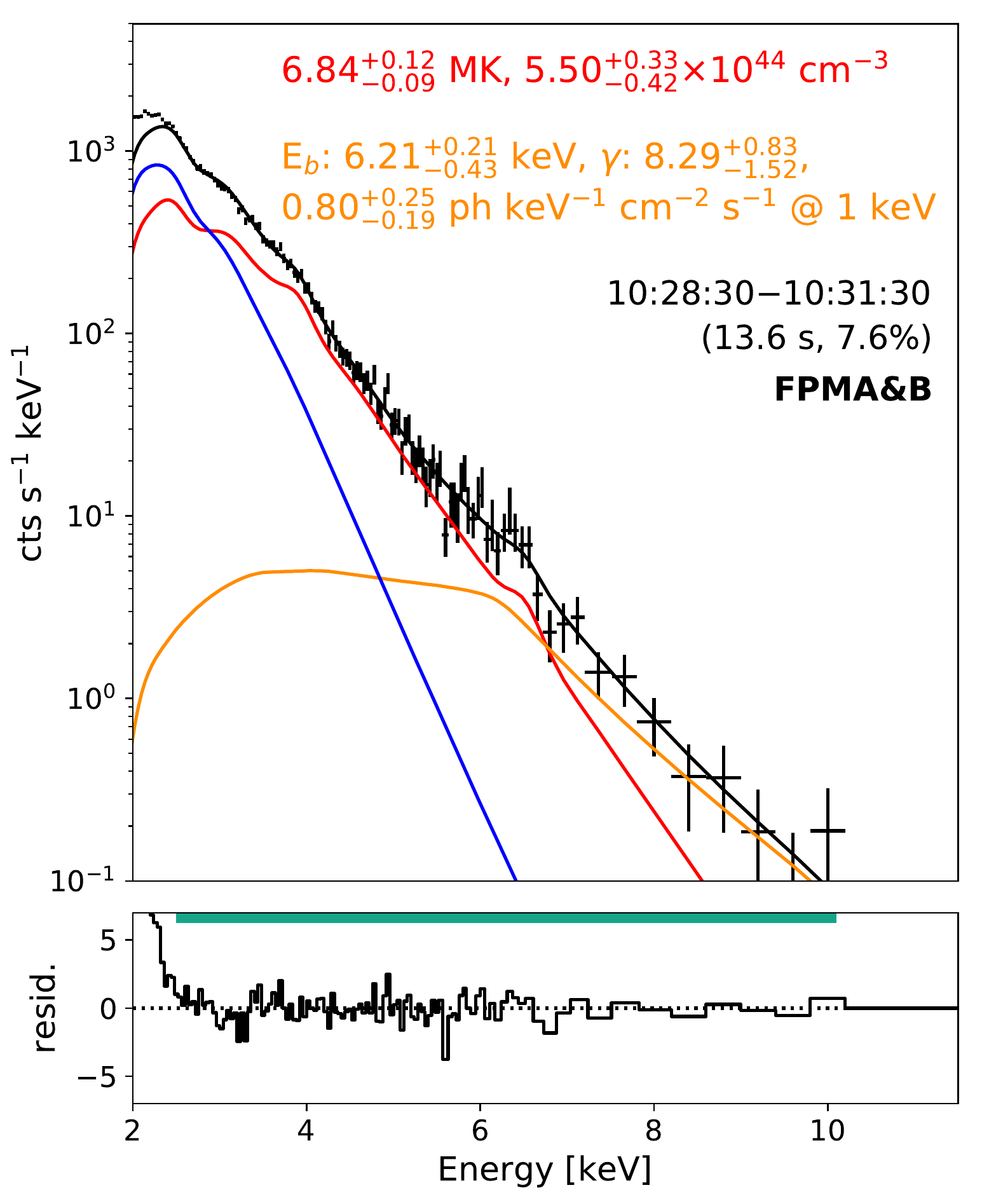}
    \caption{Microflare~3 impulsive phase spectrum shown in Figure~\ref{fig:m3} (10:28:30--10:31:30~UTC) fitted with a fixed pre-flare component (blue) with one thermal model (red) and an additional broken power-law model (orange) to represent emission from non-thermal electrons. The spectrum was taken over the time range indicated and was performed by fitting FPMA and B simultaneously. The effective exposure and livetime are also indicated in brackets.}
    \label{fig:m3_nonthermal}
\end{figure}

Figure~\ref{fig:m3_nonthermal} shows that an APEC thermal model with temperature 6.8~MK and emission measure 5.5$\times$10$^{44}$~cm$^{-3}$ in addition to a broken power-law model with a break energy of 6.2~keV, a photon index of 8.3, and a normalisation constant of 0.8~ph~keV$^{-1}$~cm$^{-2}$~s$^{-1}$ at~1~keV eliminates any excess counts above the total model (the photon index below the break was fixed at 2). The photon power-law model would provide a power of 7.03$^{+3.67}_{-2.32}\times$10$^{24}$~erg~s$^{-1}$, releasing 1.27$^{+0.66}_{-0.42}\times$10$^{27}$~erg over the 3~minute period \citep[see Equation~4 and~6 of][]{hannah_rhessi_2008}. We also find better \ion{Fe}{xviii} agreement during the rise phase with the 6.8~MK plasma predicting $\sim$57\% of the EUV emission.

The volume (V) of microflare~3 is calculated from the area (A) of averaged \ion{Fe}{xviii} emission (Figure~\ref{fig:m3}, top left panel) through the relation $V=A^{3/2}$ \citep[see][]{hannah_joint_2019} due to the complex loop network nature. Therefore, taking the largest area to encompass the emission from every phase we find an upper limit volume of 4.9$\times$10$^{27}$~cm$^{3}$. 

Performing a spectral fit over the full microflare time (10:28:30--10:37:30~UTC), with the rise time power-law and pre-flare models fixed, we find the data is fit well with a thermal model at 6.5~MK and emission measure 11.6$\times$10$^{44}$~cm$^{-3}$. Therefore, the averaged NuSTAR thermal energy for microflare~3 is 6.50$^{+0.04}_{-0.04}\times$10$^{27}$~erg and is calculated to be a A0.1 GOES class equivalent with the peak time's temperature and emission measure. This would suggest that microflare~3 is one of the weakest non-thermal microflare in current literature. All of microflare~3's spectral fit parameters are displayed in Table~\ref{tab:fits 3}.

\begin{figure*}
	\centering
	\includegraphics[width=2\columnwidth]{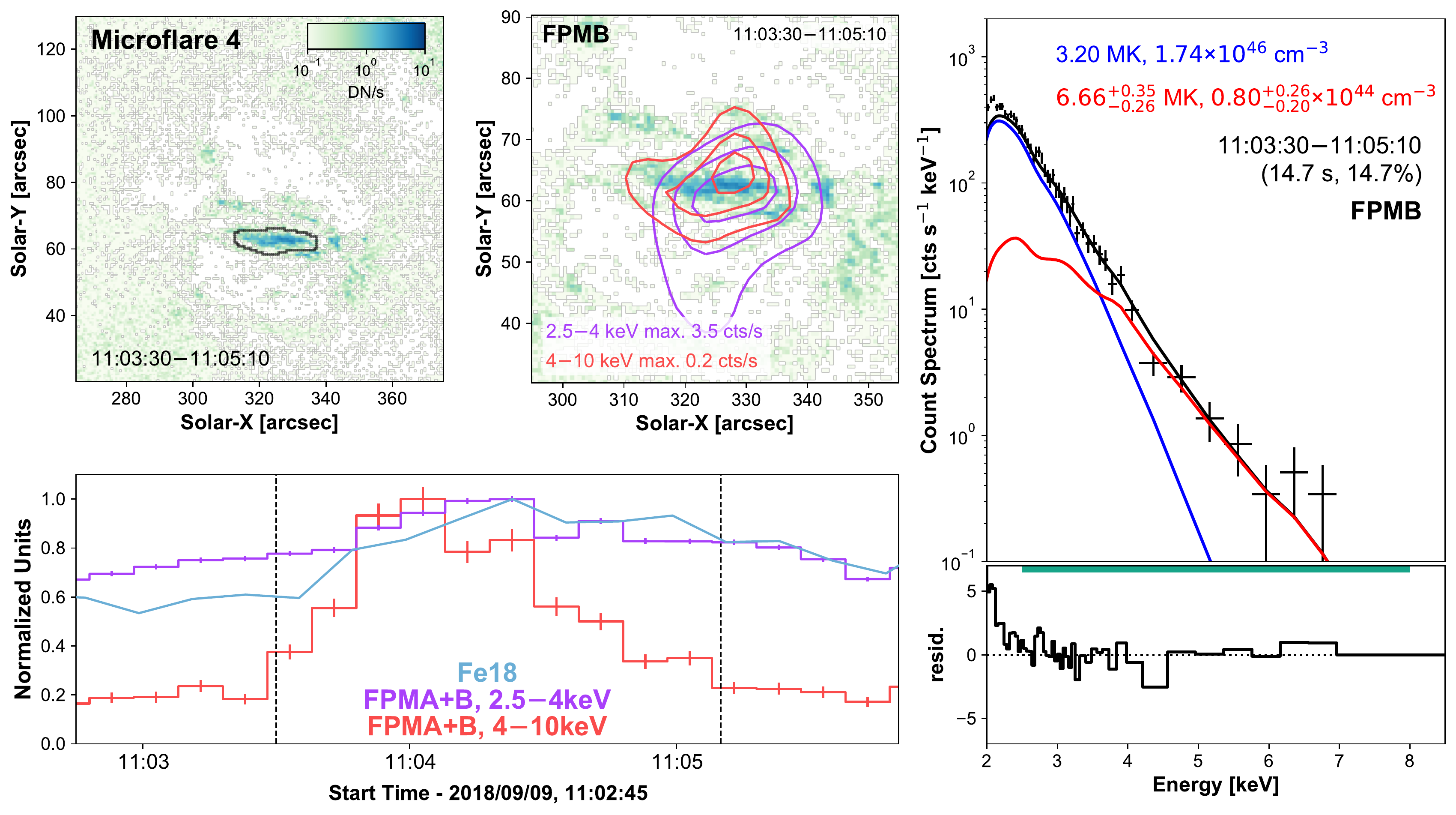}
    \caption{SDO/AIA \ion{Fe}{xviii} image integrated over microflare~4's time (top left panel). NuSTAR contours, created over the microflare time, of 2.5--4~keV (purple) and 4--10~keV (red) from FPMB are overlain on the average \ion{Fe}{xviii} emission (top middle panel) and show the 20, 50, and 80\% levels. Time profiles of \ion{Fe}{xviii} (blue) and the two NuSTAR FPMA+B energy ranges (purple, red) are plotted (bottom left panel) with the microflare time identified between two vertical dashed lines. The area used to produce the \ion{Fe}{xviii} lightcurve is shown in the top left panel with black contours, X-ray lightcurves are integrated over the full AR with 10~s binning. The X-ray spectral fit for FPMB's emission over microflare~4's time is displayed in the right panel \citep[for a more detailed analysis of microflare~4 see][]{cooper_nustar_2020}.}
    \label{fig:m4}
\end{figure*}

Occurring at 11:04~UTC in the same NuSTAR orbit, microflare~4 is the weakest X-ray microflare currently in literature and is the topic of \cite{cooper_nustar_2020} (Figure~\ref{fig:m4}). The event became more apparent when investigating higher energy ranges (4--10~keV) and is also present in the \ion{Fe}{xviii} proxy. This event benefits from having the highest livetime fraction of these data and being the only one present during this time in NuSTAR's FOV with a temperature of 6.7~MK and an emission measure of 8.0$\times$10$^{43}$~cm$^{-3}$.

From the microflare averaged \ion{Fe}{xviii} emission (Figure~\ref{fig:m4}, top left) and modelling the observed loop as a half-torus shape we find a volume 1.9$\times$10$^{25}$~cm$^{3}$. Combining this with the microflare excess temperature and emission measure we find that microflare~4 has an instantaneous thermal energy of 1.08$^{+0.23}_{-0.16}\times$10$^{26}$~erg and is an $\sim$A0.005 equivalent GOES class event. See \cite{cooper_nustar_2020} for a more in-depth analysis.

\subsection{Orbit 3: Microflare 5, 6, and 7} \label{sec:orbit 3}
\begin{figure*}
	\centering
	\includegraphics[width=2\columnwidth]{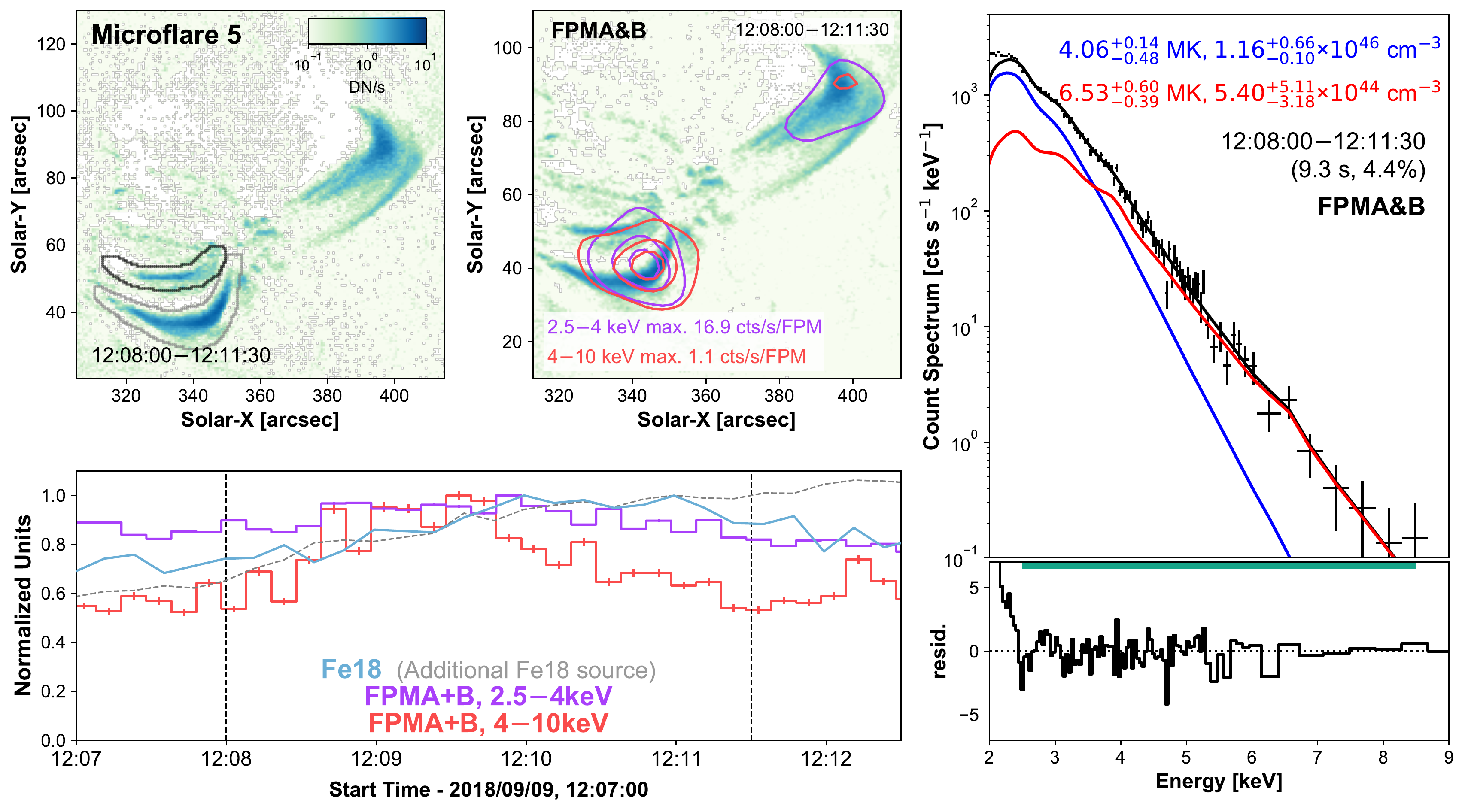}
    \caption{Same format as Figure~\ref{fig:m4} but for microflare~5. Both FPMs are able to be used in the construction of the X-ray contours and the spectral fitting. The X-ray lightcurves (bottom left panel) and spectrum (right panel) does not include emission from the large decaying loop in the top right of the \ion{Fe}{xviii} images (top left and middle panels). An additional, brighter \ion{Fe}{xviii} source is indicated in grey (top left panel) with the corresponding lightcurve shown with a grey dashed line (bottom left panel).}
    \label{fig:m5}
\end{figure*}

\begin{figure*}
	\centering
	\includegraphics[width=2\columnwidth]{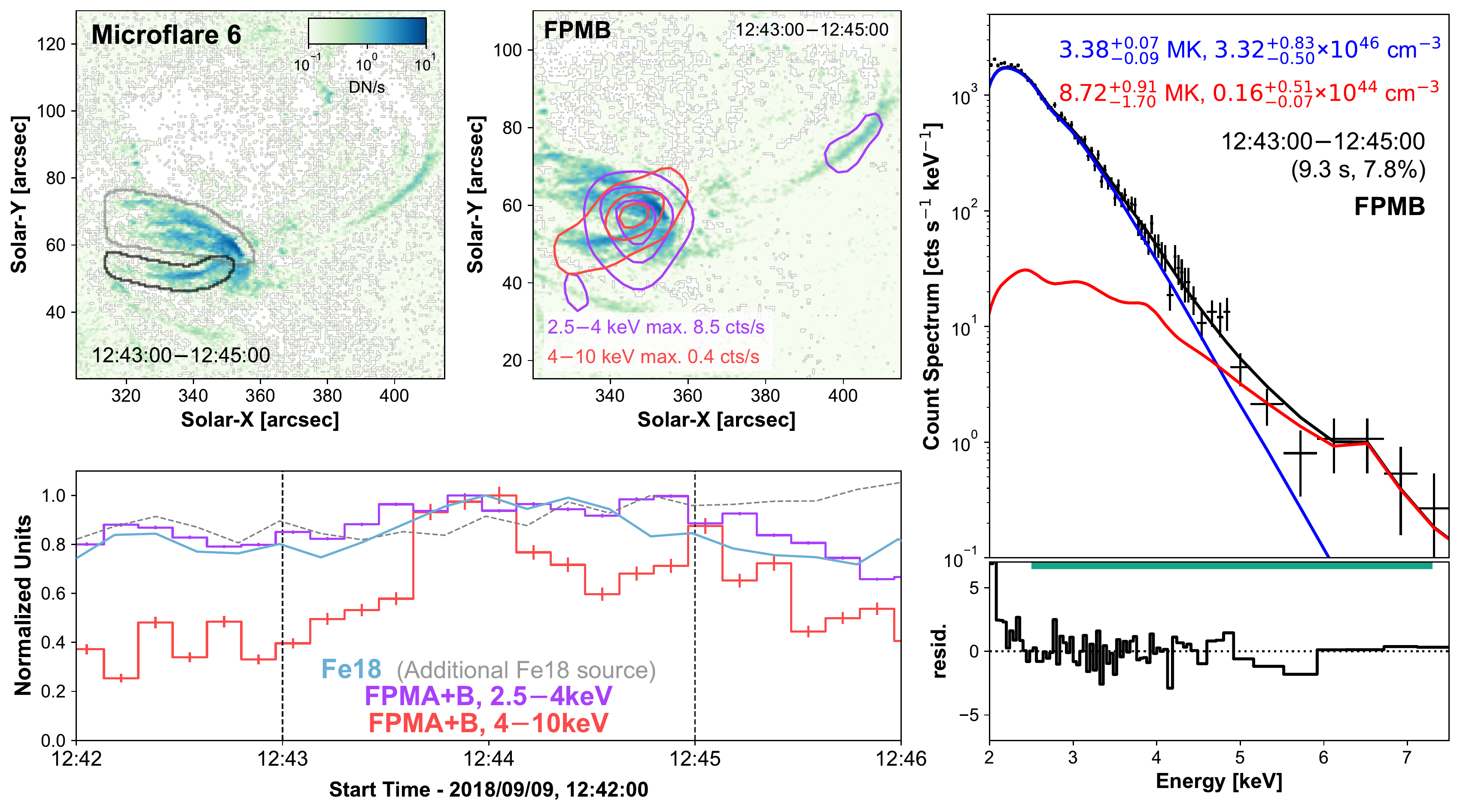}
    \caption{Same format as Figure~\ref{fig:m4} but for microflare~6. Similar to Figure~\ref{fig:m5}, an additional \ion{Fe}{xviii} source is indicated for analysis.}
    \label{fig:m6}
\end{figure*}

\begin{figure*}
	\centering
	\includegraphics[width=2\columnwidth]{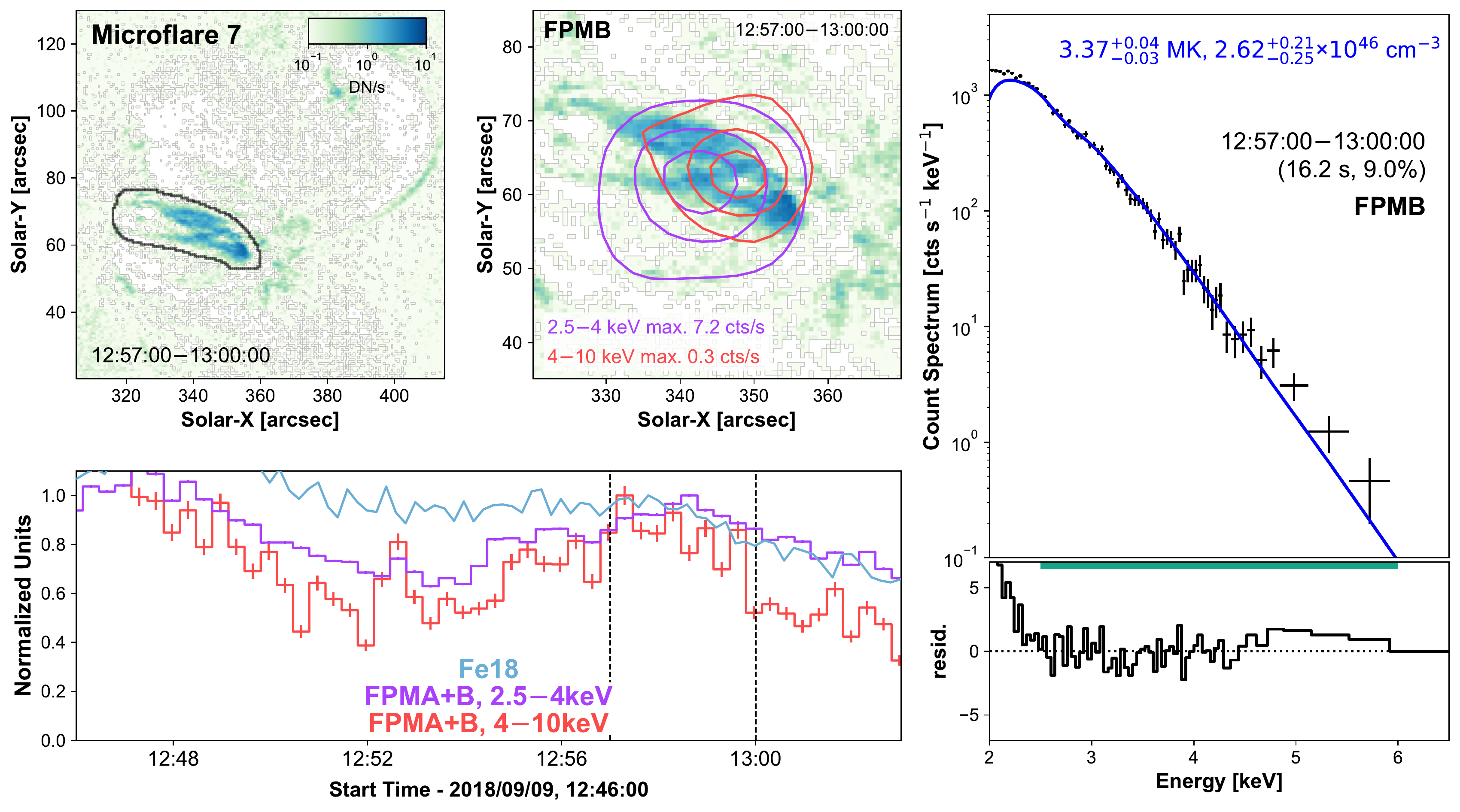}
    \caption{Same format as Figure~\ref{fig:m4} but for microflare~7 with the NuSTAR time profiles (bottom left, purple and red) binned in 20~s intervals. The elevated emission from a microflare missed due to a CHU change is clear at $\sim$12:47~UTC.}
    \label{fig:m7}
\end{figure*}

Microflare~5,~6, and~7 occur during the third and last NuSTAR dwell on September~9 and are all located around the core of the AR. They all occur with elevated background emission due to the decay of a relatively large microflare that peaked between orbit~2 and~3 at $\sim$11:45~UTC (Figure~\ref{fig:lightcurves_9th10th}, top panel) that appears to have one footpoint anchored North-West of the AR core region. The cooling loops from this larger microflare can be seen in microflare~5's \ion{Fe}{xviii} panel (and microflare~6 and~7's panels to a lesser extent) in Figure~\ref{fig:m5},~\ref{fig:m6}, and~\ref{fig:m7} (top left panels). This decay is also visible with the NuSTAR contours (Figure~\ref{fig:m5}, top middle panel). The large spatial separation of the decaying loop  allows it to be removed for time profiles (bottom left panel) and spectral fitting (right panel) via region selection. This only made the microflare clearer within its time profile with little effect on the spectral fit parameters. Both FPMA\&B were able to be used in the analysis for microflare~5.

Microflare~5 appears to be a simple loop structure, indicated with black contours in Figure~\ref{fig:m5} (top left panel). This structure is identified to be the source of raised X-rays emission for similar reasons discussed in Section~\ref{sec:orbit 1} for microflare~1 and~2. By investigating the \ion{Fe}{xviii} time profiles from both identifiable loops we find that the brighter loop displays monotonically increasing emission whereas the EUV emission from the fainter loop shows a similar profile while reaching a maximum 2~seconds after the 2.5--4~keV peak.

Comparing the X-ray microflare time profiles to the different loop structures seen in \ion{Fe}{xviii}, we identify microflare~6 coming from a similar location to microflare~5, highlighted with the black contour in Figure~\ref{fig:m6}. Microflare~7, however, comes from the loop structures North of microflare~5 and~6's location, again identified with a black contour in Figure~\ref{fig:m7}.

The orientation of the 4--10~keV NuSTAR contours also provides some corroboratory evidence that the correct microflaring loop is identified (Figure~\ref{fig:m6}, top middle panel) although this might suggest that this loop is heated more at its Western footpoint than its Eastern one. Microflare~6 is especially difficult to analyse as it occurs just before a CHU change that moved the main source of emission. The source then spent several minutes in the chip gap for both FPMs (12:45:50--12:53:21). This means that the decay of the event cannot be analysed. After microflare~6 has occurred, the general loop structure responsible for it does not produce any other event visible in \ion{Fe}{xviii}.

Microflare~7 is a small but noticeable jump in X-ray intensity (Figure~\ref{fig:m7}, bottom left panel) and appears to originate at the base of the arc feature seen clearly in \ion{Fe}{xviii} and in X-rays (Figure~\ref{fig:m7}, top left and middle panel). Microflare~7 shows evidence of higher energy X-ray emission (4--10~keV) towards the Western footpoint of the loop seen in 2.5--4~keV and \ion{Fe}{xviii} emission. This same structure appeared to undergo energy release at $\sim$12:47~UTC as seen in Figure~\ref{fig:m7} (bottom left panel); however, the same CHU change that disrupted microflare~6 caused this to be missed.

Microflare~5,~6, and~7 all show co-temporal corresponding signatures in the SDO/AIA 131~\AA{} and 171~\AA{} channels to the evolution shown in Figures~\ref{fig:m5},~\ref{fig:m6}, and~\ref{fig:m7} (bottom left panels). The structures visible in these EUV channels that are sensitive to lower temperatures (<4~MK) appear to be in the same location as the heated loops identified in \ion{Fe}{xviii} (top left panels) but are considerably smaller in size. The transient nature of these lower temperature features reveal the dynamic and multi-thermal nature of these locations, further corroborating the selection for the loops of the impulsive X-ray emission.

Despite being weak events, the X-ray spectra of both microflares~5 and~6 show a hot excess component, 6.5 and 8.7~MK respectively (Figures~\ref{fig:m5} and~\ref{fig:m6}). The X-ray spectra of microflare~7 is dominated by a single isothermal component (Figures~\ref{fig:m7}), with temperature consistent with the quiescent AR, but has a hint of more emission $>$5~keV. Unfortunately, due to the weak nature of the excess, the fitted parameters are not well constrained with the addition of another model making a fit involving two model components difficult to interpret.

By modelling the loops as half-tori we find  microflare~5 and~6 have a similar volume of 7.9$\times$10$^{25}$~cm$^{3}$. Microflare~7 is modelled with two half-tori finding a volume of 5.2$\times$10$^{26}$~cm$^{3}$. Combining these with the relevant spectral model parameters (Figure~\ref{fig:m5},~\ref{fig:m6}, and~\ref{fig:m7}, right panels) we obtain instantaneous thermal energies of 5.59$^{+1.75}_{-1.67}\times$10$^{26}$~erg, 1.30$^{+0.83}_{-0.20}\times$10$^{26}$~erg, and 5.15$^{+0.16}_{-0.20}\times$10$^{27}$~erg with equivalent GOES classifications A0.03, A0.002, and A0.2.

\subsection{Orbit 4: Microflare 8 and 9} \label{sec:orbit 4}

\begin{figure*}
	\centering
	\includegraphics[width=2\columnwidth]{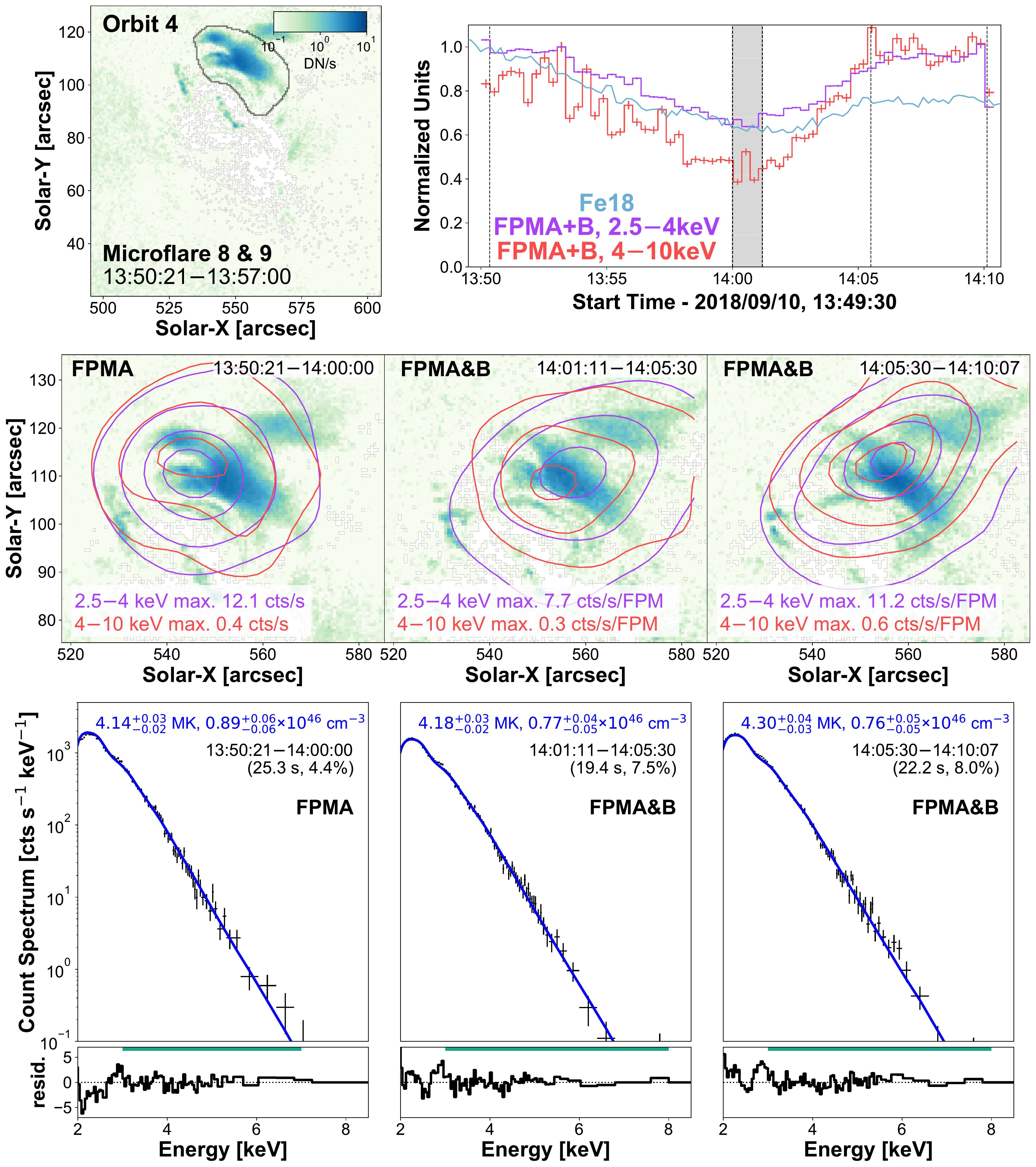}
    \caption{SDO/AIA \ion{Fe}{xviii} image integrated over microflare~8's time (top left panel) with black contours identifying the loops being investigated and the region used to produce the EUV lightcurve. Maximum normalised time profiles from \ion{Fe}{xviii} (blue) and two NuSTAR FPMA+B energy ranges (2.5--4~keV: purple, 4--10~keV: red, 20~s binning) are plotted (top right panel). The NuSTAR lightcurves are integrated over the full AR. The dark grey region indicates a time that is not used due to a noticeable shift in source location from a CHU state change. NuSTAR contours (middle row) are created over the time shown at the top right of each panel and indicated by the vertical dashed lines in the lightcurve panel. The FPMB 2.5--4~keV and 4--10~keV emission is overlain on the average \ion{Fe}{xviii} image and show the 5, 20, 50, and 80\% of the maximum emission across all time ranges for each energy range (2.5--4~keV (purple): 12.1~counts~s$^{-1}$~FPM$^{-1}$, 4--10~keV (red): 0.6~counts~s$^{-1}$~FPM$^{-1}$). The X-ray spectra over the same time ranges are shown directly below in the bottom row with the fitting range indicated by the horizontal green line above the residuals.}
    \label{fig:m8and9}
\end{figure*}

NuSTAR's first orbit on September~10 provides two examples of heated loops approximately one hour after a microflare that was observed by SDO/AIA but missed by NuSTAR. Microflare~8, therefore, shows the decay of that microflare while microflare~9 is the repeated heating of the post-flare arcade (Figure~\ref{fig:m8and9}).

Microflare~8 and~9's X-ray emission shows the tops of the heated loop arcade with the \ion{Fe}{xviii} emission also showing the loop footpoints (Figure~\ref{fig:m8and9}, top left and middle panels). There does not appear to be any spatially dynamic evolution in \ion{Fe}{xviii}. The decay of microflare~9 is missed due to NuSTAR's SAA passage and since the source is sufficiently clear of the chip-gap both FPMs are able to be used in spectral fitting and contour creation. 

The spectra for the decaying microflare~8 and the rise and peak of microflare~9 (Figure~\ref{fig:m8and9}, bottom row) are modelled well with one thermal model despite microflare~9 showing a clear microflare time profile where an excess may be expected. The decaying post flare loops of microflare~8 have a temperature of 4.1~MK and an emission measure of 8.9$\times$10$^{46}$~cm$^{-3}$ which are then slightly heated to 4.3~MK with an emission measure of 7.6$\times$10$^{46}$~cm$^{-3}$ during microflare~9; again, finding loops heated to hot AR core temperatures.

Using the same method described for microflare~3, we estimate the volume of the \ion{Fe}{xviii} loop-top source shown in Figure~\ref{fig:m8and9} (top left panel) to be 4.4$\times$10$^{27}$~cm$^{3}$ for microflare~8 and~9. We then find that the decaying loops of microflare~8 still has an equivalent GOES class of A0.1, comparable to that of the largest microflares in this study. Microflare~9 is calculated to have an energy release and GOES class  of 1.03$^{+0.02}_{-0.03}\times$10$^{28}$~erg and A0.1 at peak time.

\subsection{Orbit 5: Microflare 10} \label{sec:orbit 5}
\begin{figure*}
	\centering
	\includegraphics[width=2\columnwidth]{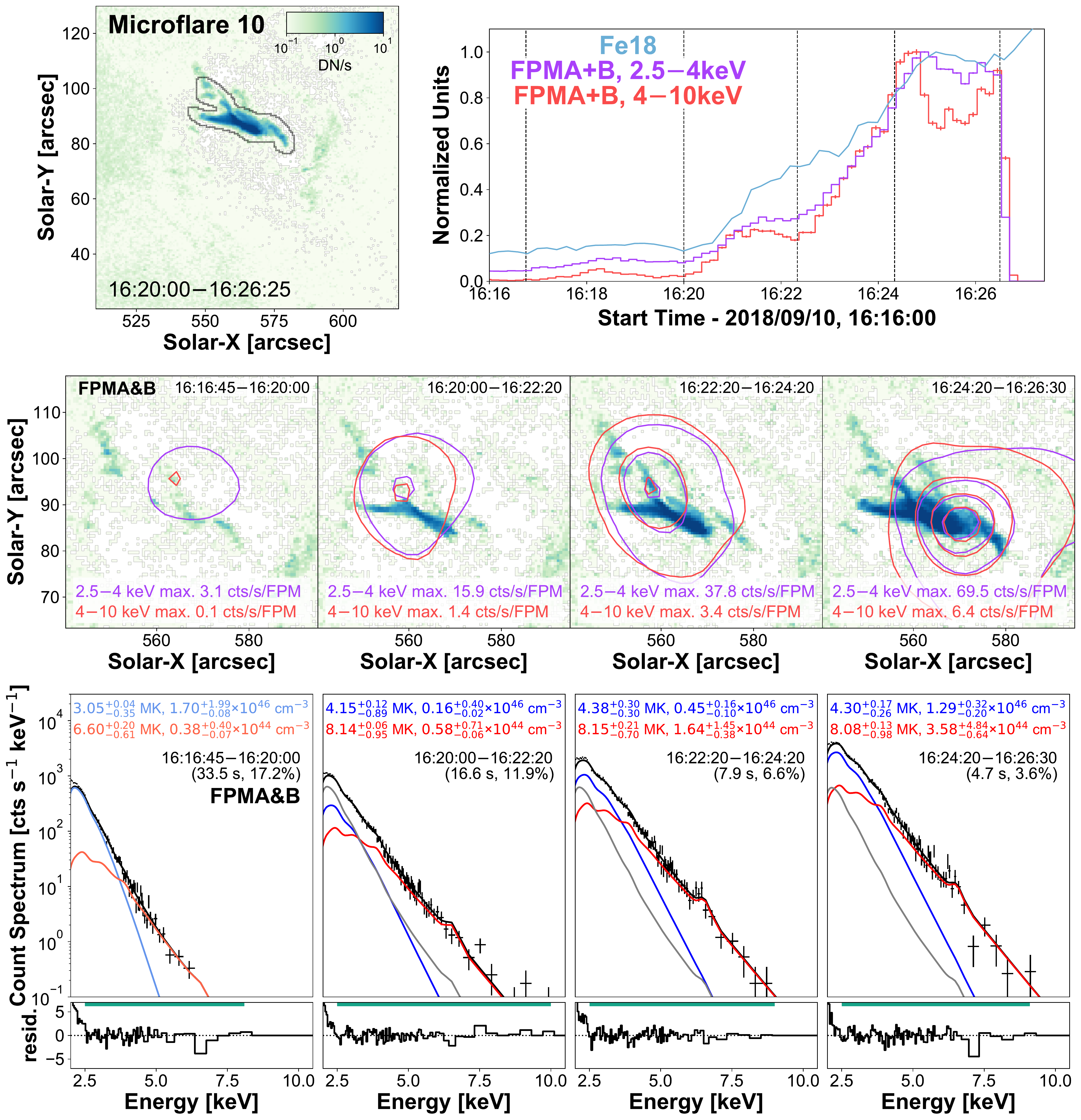}
    \caption{Same format as Figure~\ref{fig:m3} but for microflare 10. Both thermal model fits for the pre-flare stage (bottom left panel) are kept as fixed components for the other fits shown in grey. Contour levels are 2, 20, 50, and 80\% of the maximum emission across all time ranges for each energy range (2.5--4~keV (purple): 69.5~counts~s$^{-1}$~FPM$^{-1}$, 4--10~keV (red): 6.4~counts~s$^{-1}$~FPM$^{-1}$).}
    \label{fig:m10}
\end{figure*}

Microflare~10 is the brightest X-ray microflare recorded by NuSTAR from AR12721 despite going into nighttime before the peak of the event as comparison with the \ion{Fe}{xviii} lightcurve suggests (Figure~\ref{fig:lightcurves_9th10th}, panel~g and~h, and Figure~\ref{fig:m10}, top right panel). This microflare, much like microflare~3 (Figure~\ref{fig:m3}), is spatially complex and appears to be a combination of several microflaring loops (Figure~\ref{fig:m10}, top left panel). 

Pre-flare (16:16:45--16:20:00~UTC), initial rise (16:20:00--16:22:20~UTC), continued rise (16:22:20--16:24:20~UTC), and plateau (16:24:20--16:26:30~UTC) times are defined for microflare~10 and are indicated by the vertical dashed lines in the top right panel of Figure~\ref{fig:m10}. Similar to microflare~3, the contours created from the pre-flare time did not appear to localise themselves to a corresponding \ion{Fe}{xviii} source (Figure~\ref{fig:m10}, middle row, left panel). However, during the microflare the FPMA\&B X-rays contours show a small shift from the top left to the bottom right which agrees with brightening of loops seen in \ion{Fe}{xviii}.

Unlike microflare~3, the pre-flare time appears to be multi-thermal since it is fitted well with two thermal models (Figure~\ref{fig:m10}, bottom row, left panel). These pre-flare models are kept as fixed components in the other spectral fits for consistency (Figure~\ref{fig:m10}, bottom row, right three panels, grey); however, the 6.6~MK component has a negligible effect on any derived parameters. The spectral evolution after the pre-flare time indicates that the microflaring plasma is kept heated at $\sim$8~MK while increasing the amount of material. The material at $\sim$4~MK also undergoes an enhancement in emission measure and also does not vary much in temperature. 

Using the excess model parameters we can probe the consistency between the emission seen by NuSTAR to the excess observed in the SDO/AIA \ion{Fe}{xviii} synthetic flux channel. We find disagreement during all times from the microflare excess models (NuSTAR sees $\sim$112\% at the initial rise time, $\sim$128\% during the continued rise, and $\sim$135\% at the plateau). This disagreement could indicate the presence of non-thermal emission, that these models are not consistent with the \ion{Fe}{xviii} proxy, or could be because the 4~MK plasma component is at the edge of the channel's temperature response and contributes the majority of the total synthetic flux value making the comparison between NuSTAR and the composite SDO/AIA channel more notably uncertain.

Other than the inconsistency in the synthetic flux comparison, little suggests the need to introduce a non-thermal component to the spectral fitting for microflare~10 as the fits presented in Figure~\ref{fig:m10} (bottom row) appear to fit the spectra well. However, by incorporating a power-law model into the fitting we find that this would produce peak temperatures of 5.2~MK. Therefore, we find that adding a non-thermal component is not physical as microflare~10 is the brightest microflare in these data and reaches higher temperatures in its pre-flare phase (6.6~MK). In addition, including a power-law model in the spectral fitting does not resolve the disagreement in the synthetic flux comparison.

However, a non-thermal component may be expected in microflare~10 as it is as bright as microflare~3 while only being in its impulsive phase. This could be due to microflare~10 having a more complicated physical evolution with various microflaring loops of different sizes and orientation heating at different times. The more complicated evolution of microflare~10 could make the detection of any non-thermal emission difficult with it being hidden by many thermal components.

Approximating microflare~10's volumes in the same manner as microflare~3 we find upper limits of 9.9$\times$10$^{26}$, to 1.4$\times$10$^{27}$~cm$^{3}$, then to 3.6$\times$10$^{27}$~cm$^{3}$ from the initial rise, to the peak, then to the plateau. Since microflare~10's excess requires two thermal models to represent the observed emission we calculate the multi-thermal energy release ($E_{th}$) using

\begin{equation} \label{eq:multitherm}
 E_{th} = 3 k_{B} V^{\frac{1}{2}} \sum_{i=1}^{N}T_{i}EM^{1/2}_{i} \hspace{0.5cm}\text{[erg],}
\end{equation}

\noindent
where $T_{i}$ and $EM_{i}$ are the temperature and emission measure for model $i$, $V$ is the loop volume, $k_{B}$ is Boltzmann's constant, and $N$ is the total number of thermal models \citep{aschwanden_global_2015}. Therefore, using Equation~\ref{eq:multitherm}, the instantaneous thermal energy is 2.98$^{+1.25}_{-0.13}\times$10$^{27}$~erg at the initial rise and 6.08$^{+0.69}_{-0.43}\times$10$^{27}$~erg during the continued rise. At the plateau, microflare~10 achieves a GOES class equivalent of A0.3 and a thermal energy of 1.59$^{+0.19}_{-0.08}\times$10$^{28}$~erg. Therefore, microflare~10 produces the largest energy release from these data while still only being in its impulsive phase. Microflare~10's spectral fit values are summarised in Table~\ref{tab:fits 10}.

\section{A magnetic perspective} \label{sec:hmi}

The microflares shown in Section~\ref{sec:temporal and spatial and spectral} originate from sets of loops in slightly different locations but with footpoints in similar locations. All labelled events appear to be East-West orientated and rooted in two large oppositely polarised regions.

\begin{figure*}
	\centering
	\includegraphics[width=2.08\columnwidth]{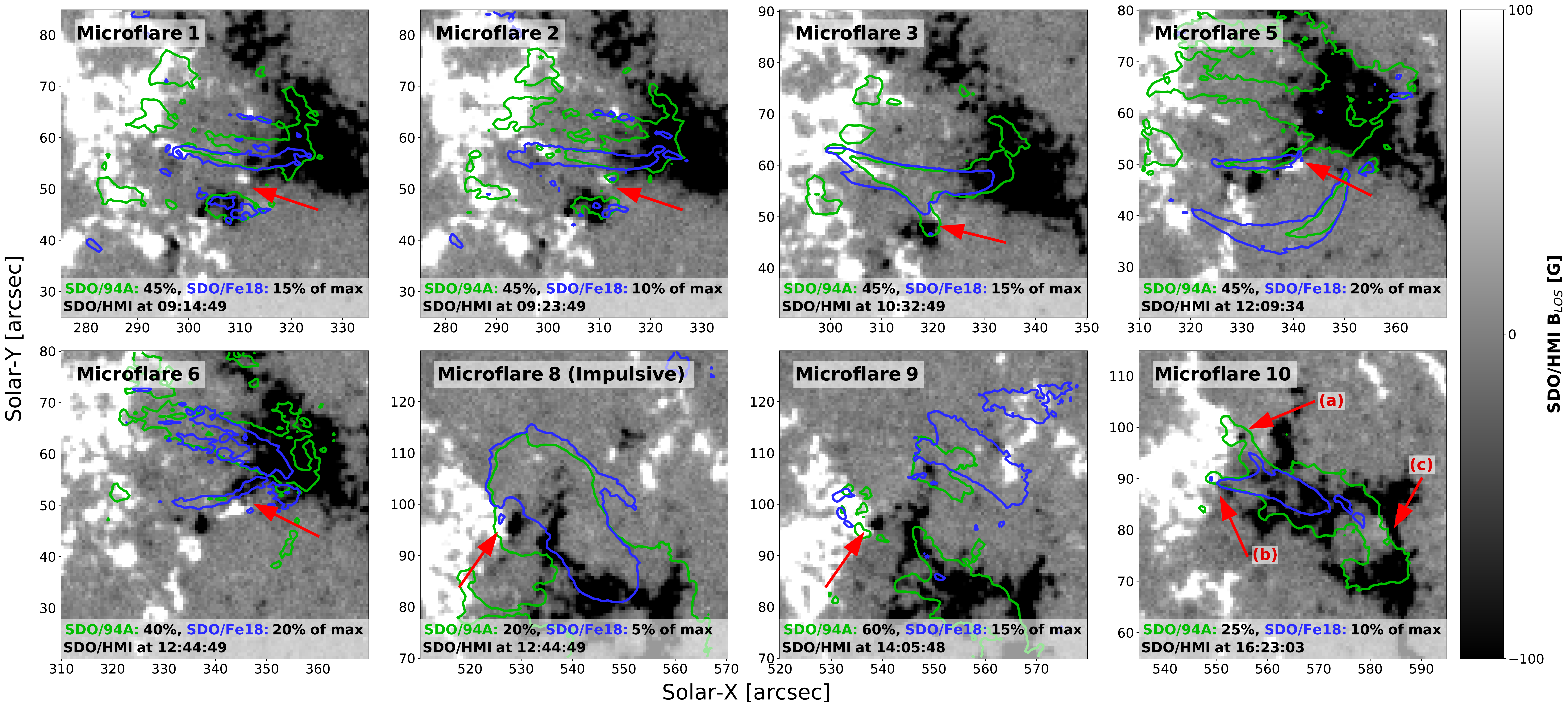}
    \caption{The SDO/HMI magnetograms at the start of the microflares with SDO/AIA 94~\AA{} (green) and the hotter \ion{Fe}{xviii} component (blue) contours at appropriate levels to observe brightening loop structures/footpoints. Red arrows indicate mixed magnetic polarity regions close to the footpoints of the identified events' coronal loops. Black and white indicates negative and positive polarity, respectively. The time used for microflare~8 covers the impulsive phase of the event (September~10, 12:40--12:50~UTC) that was not observed by NuSTAR. The time intervals used for the other microflares are those that have been defined previously.}
    \label{fig:hmi_closeup}
\end{figure*}

Most of the 10 microflares appear to have corresponding activity in the line-of-sight magnetic field at the photosphere from visual inspection of HMI images (Figure~\ref{fig:hmi_closeup}). In 8 of the 10 events, the footpoints of the loops appear to be anchored in large unipolar regions with at least one of the footpoints in close proximity to a smaller oppositely polarised parasitic region or overlying two oppositely polarised patches. Visual inspection of SDO/AIA 1600~\AA{} and 1700~\AA{} shows bright chromospheric material at the footpoint locations suggested by SDO/AIA 94~\AA{} and \ion{Fe}{xviii} in Figure~\ref{fig:hmi_closeup}.

Figure~\ref{fig:hmi_closeup} shows potential flux cancellation or emerging regions with red arrows for 8 microflares on SDO/HMI magnetograms at the microflare start time. The SDO/AIA 94~\AA{} and hotter \ion{Fe}{xviii} emission is shown with green and blue contours, respectively, at levels that best show the loops in question. Red arrows indicate the mixed polarity regions close to or at the footpoints of microflare~1,~2,~3,~5,~6,~8,~9, and~10. Microflare~10 has prominent mixed polarity regions at three footpoints and so are labelled~a,~b, and~c.

The microflares' properties---such as temperatures, emission measures, and strong \ion{Fe}{xviii} presence---are similar to previously studied microflares that have evidence of photospheric magnetic flux cancellation \citep{chitta_solar_2017, chitta_compact_2017, chitta_nature_2018, chitta_energetics_2019}. This corroborates studies that suggest \ion{Fe}{xviii} intensity in loops and ARs is correlated to the presence of magnetic flux emergence or cancellation \citep{asgari-targhi_study_2019,chitta_impulsive_2020}.

Flux cancellation at low levels in the atmosphere may be the mechanism by which multiple microflaring events can appear to take place in the same general loop structures \citep{chitta_impulsive_2020}. Microflare~1 and~2 appear to occur in similarly positioned loops while a newly emerged positive polarity migrates South, under the flaring loops. Microflare~5 and~6 also seem to occur in a similar loop and, therefore, the dynamic positive polarity region close to the Western footpoint may have been the trigger. Similarly, the negative polarity at the Eastern footpoint could also explain microflare~9 taking place in the same overall structure as the microflare that lead to the decay identified as Event~8 (Figure~\ref{fig:hmi_closeup}).

Although there does not appear to be any visible ``parasitic'' polarity at the footpoints of microflare~7, the asymmetric loop heating (Figure~\ref{fig:m7}, top left and middle panels) could be explained by flux emergence or cancellation at only the Western loop footpoint due to an  unresolved opposite polarity area. It has been shown that with instruments with greater spatial resolution than HMI that apparent unipolar areas can actually have opposite polarity regions \citep{chitta_solar_2017}. Single footpoint magnetic flux cancellation could be the reason a loop appears to be asymmetrically heated \citep{testa_coronal_2020}. Microflare~6 may have undergone a similar scenario.

Only the magnetic configuration at the two largest events' footpoints are investigated further as these parasitic polarities are easily isolated from regions of the same polarity and appear to only interact with one oppositely polarised region (Figure~\ref{fig:hmi_closeup}, microflares~3 and~10). Even though the other microflares provide compelling visual evidence of intriguing magnetic activity it is beyond the scope of this study to analyse these events quantitatively.

\begin{figure}
	\centering
	\includegraphics[width=\columnwidth]{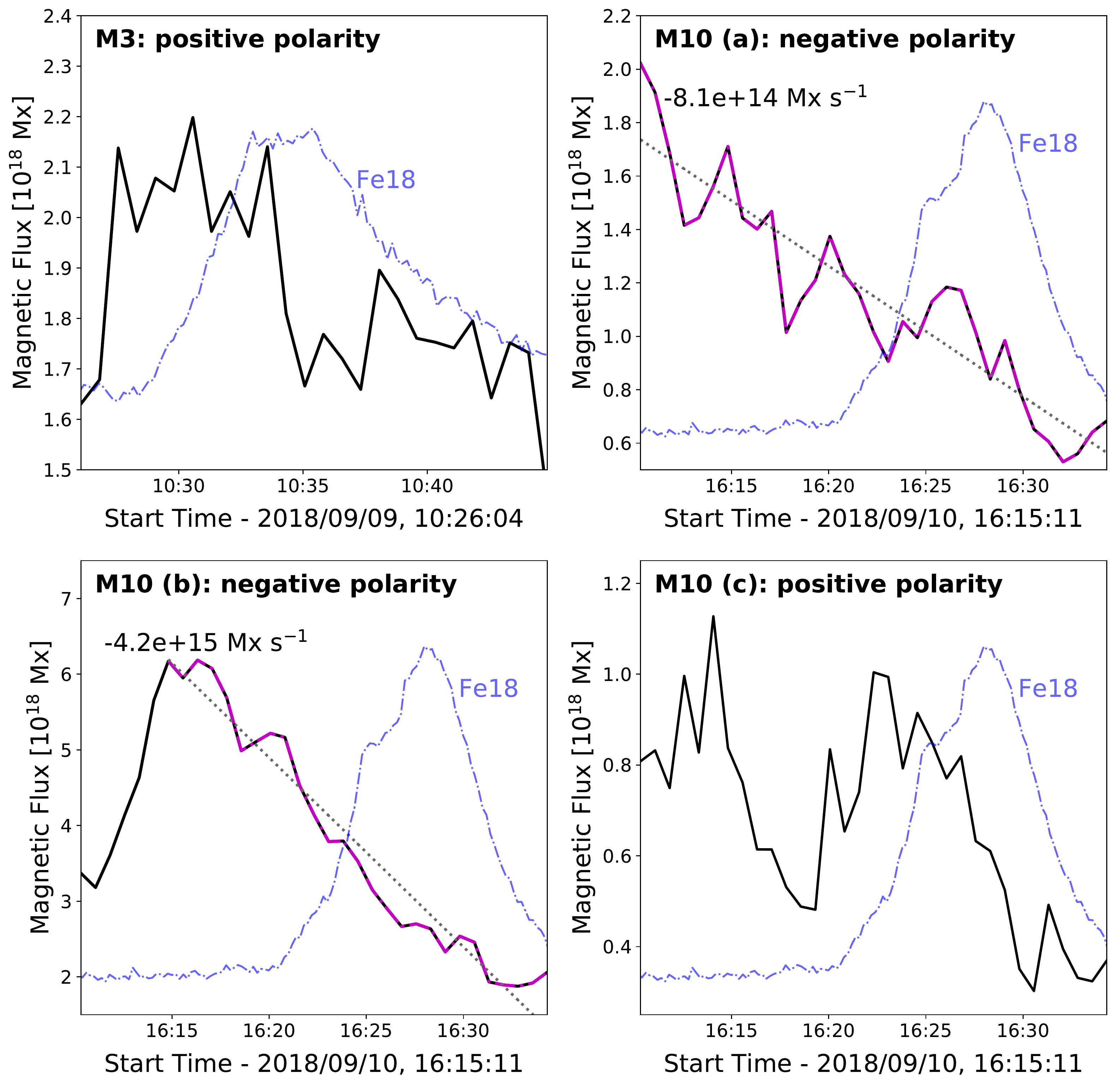}
    \caption{Evolution of the parasitic magnetic polarity for microflare~3 (top left panel) and~10 (top right and bottom panels) identified in Figure~\ref{fig:hmi_closeup}. A straight line was fitted where steady magnetic flux cancellation was observed (grey, dotted) with the cancellation rate displayed in Mx~s$^{-1}$. The linear fit took place over the times indicated by the magenta dashdotted line. The AR \ion{Fe}{xviii} evolution for microflare~3 and~10 is shown (blue, dashdotted).}
    \label{fig:m3m10footpoints}
\end{figure}

Figure~\ref{fig:m3m10footpoints} shows the photospheric magnetic flux evolution of each identified parasitic polarity at the footpoints of microflare~3 (M3: one positive parasitic polarity) and microflare~10 where two negative polarities have been identified (M10 (a) and (b)) and one positive polarity (M10 (c)). Both parasitic positive polarities show flux emergence then cancellation during the onset of their respective flares (Figure~\ref{fig:m3m10footpoints}, top left and bottom right) while both negative polarities for microflare~10 show constant flux cancellation (Figure~\ref{fig:m3m10footpoints}, top right and bottom left). Care was taken to ensure that no other magnetic flux with the same polarity as the parasitic feature was included in the region used to obtain the magnetic flux and only line-of-sight magnetic field strengths with magnitudes $>$17~G were used to only include pixels above noise levels \citep{pesnell_solar_2012}.

We find flux cancellation on the order of 10$^{14}$--10$^{15}$~Mx~s$^{-1}$ close to the apparent locations of microflare~10's footpoints. Monotonic magnetic flux cancellation of this order has been associated with brightening features as well as the production of A and B class microflares \citep{chitta_solar_2017,chitta_nature_2018}. Similar flux cancellation rates could liberate 10$^{27}$--10$^{28}$~erg of magnetic energy over a period of 30~minutes, depending on current sheet lengths and structure magnetic field strengths \citep{chitta_impulsive_2020}. This suggests that it could be possible for microflare~10, a sub-A class flare but with flux cancellation occurring for $>$15~minutes at footpoints~a and~b, to be easily triggered and significantly powered by this process.

\section{Discussion and Conclusions}

In this paper we present the largest study of microflares occurring in a single active region observed with NuSTAR. We significantly increase the number of investigated NuSTAR microflares in current literature helping to provide a more statistical view of flares and their nature at this scale.

The majority of the flares studied here show more impulsive, and earlier peaking time profiles at higher HXR ranges (4--10~keV) when compared to the lower energy ranges (2.5--4~keV and \ion{Fe}{xviii}), indicative of hotter material in the earlier stages of the flare or non-thermal emission, similar to properties observed in larger flares. 

All 10 microflares observed by NuSTAR presented unique challenges when attempting to isolate the microflare excess or enhanced AR emission. However, each microflare was easily identified when analysing NuSTAR's HXR data. Once a microflare had been found in X-rays it could then be investigated further in the \ion{Fe}{xviii} proxy channel. By utilising the observed corresponding behaviour of \ion{Fe}{xviii} emission to HXRs the individual microflaring loops were identified. We find good agreement between the identified loops in the \ion{Fe}{xviii} lightcurves and the NuSTAR AR integrated 2.5--4~keV time profiles. The higher X-ray energy range, 4--10~keV, generally displays a more impulsive feature. Although these incredibly weak events are identifiable with the \ion{Fe}{xviii} proxy, HXRs show microflares as more pronounced above the surrounding emission over a larger area. 

The importance of HXR data is most clear when considering microflare~1 and~2 or~5 and~6 as the brightest and most obvious \ion{Fe}{xviii} loops, indicated by cyan contours in Figure~\ref{fig:m1and2} and grey contours in Figure~\ref{fig:m5} and~\ref{fig:m6}, are determined not to be the likely sources of the X-ray microflares. In these scenarios the impulsive X-ray emission appears to be coming from much weaker, but transient, \ion{Fe}{xviii} sources. This has important implications for future studies when identifying common X-ray and EUV sources in a spatial context, especially in analysis methods that rely on image data.

We also find that the loop footpoints appear to be visibly rooted in, or close to, mixed magnetic polarity regions in the photosphere in 8/10 microflares. The possibility of flux cancellation or emergence at these footpoints could explain why there is repeated heating of the same general structure in microflares~1 to~2,~5 to~6, and~8 to~9. We also present evidence of positive magnetic flux emergence at an apparent footpoint in microflare~3 and~10 as well as constant flux cancellation of negative parasitic polarities at two other footpoints of microflare~10. \cite{chitta_impulsive_2020} showed that the presence of flux cancellation/emergence at the photosphere interacting with the overlying coronal loops may be connected to the onset of microflares in the cores of ARs and may play a significant role in driving impulsive heating. The flux cancellation rates calculated for microflare~10, 10$^{14}$--10$^{15}$~Mx~s$^{-1}$, are consistent with other findings \citep{chitta_solar_2017,chitta_nature_2018}.

We find that the brightest microflares (3 and~10) have more complicated loop configurations compared to the weaker microflares in this study. Microflare~3 also shows that the hottest temperatures are reached during the initial phase \citep{mitra-kraev_solar_2019, testa_coronal_2020}. This is difficult to conclude with microflare~10 as the peak and the decay was not recorded, although some of the hottest temperatures (8.2~MK) within these data were obtained during its initial phase. 

Microflare~3 is also found to have non-thermal emission present during its rise time injecting an energy of 1.3$\times$10$^{27}$~erg during the impulsive phase with an equivalent GOES class of A0.1 and reaching temperatures up to 7~MK. The non-thermal energy is smaller than the thermal (6.5$\times$10$^{27}$~erg), indicating that the non-thermal component is not solely responsible for all heating or that we do not observe all of the non-thermal emission. However, the method used to calculate the microflare's volume produces a conservative upper limit and by applying a different method, or introducing a filling factor, the microflare's thermal energy could be significantly reduced and, therefore, heated just by the accelerated electrons. Microflare 3 is the faintest non-thermal microflare so far observed with NuSTAR. Compared to the previously reported A5.7 microflare \citep{glesener_accelerated_2020} this event has a considerably weaker non-thermal component due to a steeper spectra and higher break energy.

In this paper, we present evidence of: (1) several GOES sub-A class equivalent microflares with energies between 10$^{26}$--10$^{28}$~erg reaching temperatures up to 10~MK, similar to temperatures linked to hot and transient coronal loops and also to the initial phases of microflares \citep{testa_iris_2020,testa_coronal_2020}; (2) an A0.1 equivalent non-thermal X-ray microflare, one of the weakest in literature; and (3) the presence of mixed polarity magnetic fields at, or close to, the footpoints of the majority of the observed X-ray microflares, which may explain the repeated microflaring of similar loop structures, with measured flux cancellation comparable to that found in other microflares \citep{chitta_impulsive_2020}.


\section*{Acknowledgements}
This paper made use of data from the NuSTAR mission, a project led by the California Institute of Technology, managed by the Jet Propulsion Laboratory, funded by the National Aeronautics and Space Administration. These observations were supported through the NuSTAR Guest Observer program (NASA grant 80NSSC18K1744). This research used version 2.0.6 \citep{stuart_j_mumford_2021_4421322} of the SunPy open source software package \citep{sunpy_community_sunpy_2020}, version 0.4.0 \citep{barnes_w_t_2020_4315741} of the aiapy open source software package \citep{barnes_aiapy_2020}, and made use of Astropy,\footnote{\url{http://www.astropy.org}} a community-developed core Python package for Astronomy \citep{2018AJ....156..123T}. Other Python packages that were extensively used were Matplotlib \citep{Hunter:2007}, NumPy \citep{harris2020array}, and SciPy \citep{2020SciPy-NMeth}. This research also made use of HEASoft (a unified release of FTOOLS and XANADU software packages) and NuSTAR Data Analysis Software (NuSTARDAS). This paper also made use of the SolarSoft IDL distribution (SSW) from the IDL Astronomy Library. We would like to thank the anonymous reviewer for their helpful feedback.

K.C. is supported by a Royal Society Research Fellows Enhancement Award and I.G.H is supported by a Royal Society University Fellowship.

\section*{Data Availability}

All data used is publicly available. SDO data can be obtained from the Joint Science Operations Center (JSOC)\footnote{\url{http://jsoc.stanford.edu/}} using SunPy's \verb|Fido|\footnote{\url{https://docs.sunpy.org/en/stable/guide/acquiring_data/fido.html\#fido-guide}} object while the NuSTAR data is available from the NuSTAR Master Catalog\footnote{\url{https://heasarc.gsfc.nasa.gov/db-perl/W3Browse/w3table.pl?tablehead=name=numaster&Action=More+Options}} with the OBSIDs 80414201001, 80414202001, 80414203001, 80415201001, 80415202001, and 80415203001.



\bibliographystyle{mnras}
\bibliography{ref} 




\appendix

\section{Spectral Fit Properties}
\begin{table*}
 \caption{APEC isothermal model fits of 8 out of 10 events (excluding the two brightest events) with associated loop volume, instantaneous thermal energy, and GOES equivalent class. The phase(s) of each microflare within the time range investigated and the suitable FPM(s) for spectral analysis are also indicated. All errors are obtained through MCMC analysis and 1-$\sigma$ equivalent.}
 \label{tab:xspec fits}
 \begin{tabular*}{1.95\columnwidth}{@{}l@{\hspace*{10pt}}l@{\hspace*{10pt}}l@{\hspace*{10pt}}l@{\hspace*{10pt}}l@{\hspace*{6pt}}l@{\hspace*{8pt}}l@{\hspace*{8pt}}l@{}}
  \hline
  Microflare & Phase(s) of & FPM & Background/Pre-flare     & Excess/Microflare               & Loop Volume          & Inst. Thermal Energy       & GOES\\
        & Microflare & Used &(T [MK], EM [$\times$10$^{46}$~cm$^{-3}$])  & (T [MK], EM [$\times$10$^{44}$~cm$^{-3}$])  & [cm$^{3}$] & [erg]  & Class    \\
  \hline
  
  1 & All & B &  3.59$^{+0.14}_{-0.20}$, 1.07$^{+0.42}_{-0.19}$  &  8.14$^{+0.40}_{-1.68}$, 0.14$^{+0.59}_{-0.05}$  &  2.2$\times$10$^{26}$ & 1.87$^{+1.51}_{-0.26}\times$10$^{26}$   &  A0.001\rule{0pt}{2.6ex}\\[5pt] 
  
  2$^a$     & Rise/Peak & B &  3.17$^{+0.07}_{-0.13}$, 2.99$^{+0.42}_{-0.71}$  &  5.17$^{+0.61}_{-0.75}$, 4.24$^{+9.37}_{-2.85}$  &   2.2$\times$10$^{26}$ & 6.53$^{+3.47}_{-2.36}\times$10$^{26}$   &  A0.01\\[5pt]
  
  4$^b$ & All & B &  3.20, 1.74  &  6.66$^{+0.69}_{-0.71}$, 0.80$^{+0.67}_{-0.32}$  &  1.9$\times$10$^{25}$ & 1.08$^{+0.23}_{-0.16}\times$10$^{26}$ &  A0.005\\[5pt] 
  
  5 & All & A\&B &  4.06$^{+0.14}_{-0.48}$, 1.16$^{+0.66}_{-0.10}$  &  6.53$^{+0.60}_{-0.39}$, 5.40$^{+5.11}_{-3.18}$  &  7.9$\times$10$^{25}$ & 5.59$^{+1.75}_{-1.67}\times$10$^{26}$ &   A0.03\\[5pt]
  
  6 & Rise/Peak & B &  3.38$^{+0.07}_{-0.09}$, 3.32$^{+0.83}_{-0.50}$  &  8.72$^{+0.91}_{-1.70}$, 0.16$^{+0.51}_{-0.07}$  &  7.9$\times$10$^{25}$ & 1.30$^{+0.83}_{-0.20}\times$10$^{26}$ &   A0.002\\[5pt]
  
  7& All & B &    &  3.37$^{+0.04}_{-0.03}$, 262.16$^{+21.36}_{-25.28}$  &  5.2$\times$10$^{26}$ & 5.15$^{+0.16}_{-0.20}\times$10$^{27}$ &  A0.2\\[5pt]
  
  8 & Decay & A &    &  4.14$^{+0.03}_{-0.02}$, 89.22$^{+5.56}_{-6.18}$  &  4.4$\times$10$^{27}$ & 1.07$^{+0.03}_{-0.03}\times$10$^{28}$ &  A0.1\\[5pt]
  
  9 & Peak & A\&B &   &  4.30$^{+0.04}_{-0.03}$, 76.38$^{+4.57}_{-5.11}$  &  4.4$\times$10$^{27}$ & 1.03$^{+0.02}_{-0.03}\times$10$^{28}$ &  A0.1\\[5pt]
  
  \hline
  \multicolumn{8}{l}{$^a$ Background component was constrained between 3--4~MK}\\
  \multicolumn{8}{l}{$^b$ Pre-flare parameters were fixed components during the microflare fitting}\\
 \end{tabular*}
\end{table*}

\begin{table*}
 \caption{Spectral fit parameters for microflare~3 where both FPMA and B were fitted simultaneously. The pre-flare model was fixed in the rise, peak, and decay times. The photon index below E$_{B}$ is fixed at 2.}
 \label{tab:fits 3}
 \begin{tabular*}{1.95\columnwidth}{@{}l@{\hspace*{9pt}}l@{\hspace*{9pt}}l@{\hspace*{7pt}}l@{\hspace*{9pt}}l@{\hspace*{9pt}}l@{\hspace*{8pt}}l@{}}
  \hline
   Phase of & Thermal Model     & Non-thermal Model      & Loop Volume          &  Inst. Thermal & Non-Thermal     & GOES\\
         Microflare &(T [MK], EM [$\times$10$^{44}$~cm$^{-3}$])  & (E$_{B}$ [keV], $\gamma$,  & [cm$^{3}$] & Energy &  Energy  & Class    \\
          &  & Norm [ph keV$^{-1}$ cm$^{-2}$ s$^{-1}$ at 1~keV])  & & [erg] &  [erg]  &     \\
  \hline
  
   Pre-flare &  4.09$^{+0.02}_{-0.02}$, 63.00$^{+3.00}_{-3.00}$ &      &   &   & & \rule{0pt}{2.6ex}\\[5pt]
  
        Rise &  6.84$^{+0.12}_{-0.09}$, 5.50$^{+0.33}_{-0.42}$ &  6.21$^{+0.21}_{-0.43}$, 8.29$^{+0.83}_{-1.52}$, 0.80$^{+0.25}_{-0.19}$    & 3.5$\times$10$^{27}$ & 3.93$^{+0.07}_{-0.09}\times$10$^{27}$   & 1.27$^{+0.66}_{-0.42}\times$10$^{27}$ &  A0.04\rule{0pt}{2.6ex}\\[5pt]
       
        Peak &  6.65$^{+0.03}_{-0.03}$, 15.97$^{+0.41}_{-0.44}$ &      & 3.5$\times$10$^{27}$ & 6.53$^{+0.06}_{-0.06}\times$10$^{27}$ & & A0.1\rule{0pt}{2.6ex}\\[5pt]
       
        Decay &  5.75$^{+0.08}_{-0.06}$, 23.65$^{+1.24}_{-1.41}$ &      & 4.9$\times$10$^{27}$ & 8.14$^{+0.12}_{-0.14}\times$10$^{27}$ & & A0.1\rule{0pt}{2.6ex}\\[5pt]
  
  \hline
   Whole Time &  6.54$^{+0.02}_{-0.02}$, 11.60$^{+0.20}_{-0.20}$ &  6.21, 8.29, 0.27$^a$    & 4.9$\times$10$^{27}$ & 6.50$^{+0.04}_{-0.04}\times$10$^{27}$ & 1.27$\times$10$^{27}$ & A0.07\rule{0pt}{2.6ex}\\
   (Rise--Decay) & & & & &  \\[5pt]
  \hline
  \multicolumn{6}{l}{$^a$ Rise time broken power-law components are fixed and the normalisation parameter is scaled to the full time interval spectrum}\\
 \end{tabular*}
\end{table*}

\begin{table*}
 \caption{Spectral fit parameters for microflare~10 where both FPMA and B were fitted simultaneously. The total pre-flare model was fixed in the rise, peak, and decay times.}
 \label{tab:fits 10}
 \begin{tabular*}{1.65\columnwidth}{@{}l@{\hspace*{10pt}}l@{\hspace*{10pt}}l@{\hspace*{10pt}}l@{\hspace*{10pt}}l@{\hspace*{10pt}}l@{\hspace*{10pt}}l@{}}
  \hline
   Phase of & Thermal Model 1     & Thermal Model 2      & Loop Volume          & Inst. Thermal Energy      & GOES\\
        Microflare &(T [MK], EM [$\times$10$^{46}$~cm$^{-3}$])  & (T [MK], EM [$\times$10$^{44}$~cm$^{-3}$])  & [cm$^{3}$] & [erg]  & Class    \\
  \hline
   Pre-flare &  3.05$^{+0.04}_{-0.35}$, 1.70$^{+1.99}_{-0.08}$  &  6.60$^{+0.20}_{-0.61}$, 0.38$^{+0.40}_{-0.07}$   &  &   &  \rule{0pt}{2.6ex}\\[5pt]
  
           Init. Rise &  4.15$^{+0.12}_{-0.89}$, 0.16$^{+0.40}_{-0.02}$  &  8.14$^{+0.06}_{-0.95}$, 0.58$^{+0.71}_{-0.06}$   &  9.9$\times$10$^{26}$ & 2.98$^{+1.25}_{-0.13}\times$10$^{27}$ &  A0.03\\[5pt]
          
           Cont. Rise &  4.38$^{+0.30}_{-0.30}$, 0.45$^{+0.16}_{-0.10}$  &  8.15$^{+0.21}_{-0.70}$, 1.64$^{+1.45}_{-0.38}$   &  1.4$\times$10$^{27}$ & 6.08$^{+0.69}_{-0.43}\times$10$^{27}$ &  A0.1\\[5pt]
          
           Plateau &  4.30$^{+0.17}_{-0.26}$, 1.29$^{+0.32}_{-0.20}$  &  8.08$^{+0.13}_{-0.98}$, 3.58$^{+4.84}_{-0.64}$  & 3.6$\times$10$^{27}$ & 1.59$^{+0.19}_{-0.08}\times$10$^{28}$ &  A0.3\\[5pt]
  
  \hline
 \end{tabular*}
\end{table*}


\bsp	
\label{lastpage}
\end{document}